\documentclass{jfm}

\usepackage{pgf}
\usepackage{pgfplots}
\pgfplotsset{width=2.4in, height=1.9in,every axis plot/.append style={thick}, compat=1.9} 
\usepackage{graphicx}	
\usepackage{epstopdf, epsfig}
\usepackage{ctable}

\graphicspath{{./fig/}}

\newcommand{\Ra}{\RA}
\newcommand{\Nu}{\NU}

\newcommand\NU{\mbox{\textit{Nu}}}
\newcommand\RA{\mbox{\textit{Ra}}}
\newcommand\Ro{\mbox{\textit{Ro}}}
\newcommand\Ek{\mbox{\textit{Ek}}}

\newcommand{\avg}[2]{\langle#1\rangle_{\mathrm{#2}}}

\renewcommand{\vec}[1]{\mathbf{#1}}

\definecolor{gfblue4}{RGB}{0, 102, 155}
\definecolor{gfblue3}{RGB}{0, 137, 204}
\definecolor{gfblue2}{RGB}{7, 170, 255}
\definecolor{gfblue1}{RGB}{130, 211, 255}
\definecolor{gfred1}{RGB}{255, 117, 186}
\definecolor{gfred2}{RGB}{255, 0, 124}
\definecolor{gfred3}{RGB}{196, 0, 96}
\definecolor{gfred4}{RGB}{153, 0, 76}

\definecolor{gfblue}{RGB}{0, 102, 156}
\definecolor{gfred}{RGB}{154, 0, 77}

\definecolor{cedit}{rgb}{.7,.2,.0}

\definecolor{cc}{rgb}{1,.0,.0}

\definecolor{cexp}{rgb}{.0,.3,.6}

\definecolor{cdns}{rgb}{.7,.2,.1}

\definecolor{cdns}{rgb}{.7,.2,.1}

\usepackage{verbatim}
\usepackage{lscape}
\usepackage{booktabs}
\usepackage{amsmath,bm}
\usepackage{placeins}

\definecolor{gfblue4}{RGB}{0, 102, 155}
\definecolor{gfred4}{RGB}{153, 0, 76}

\title{Boundary zonal flows in rapidly rotating turbulent thermal convection}

\author{Xuan Zhang\aff{1}\corresp{\email{xuan.zhang@ds.mpg.de}}, Robert E. Ecke\aff{2,3}, \and Olga Shishkina\aff{1}\corresp{\email{Olga.Shishkina@ds.mpg.de}}}

\affiliation{\aff{1}Max Planck Institute for Dynamics and Self-Organization,\\ Am Fassberg 17, 37077 G\"ottingen, Germany\\
\aff{2}Center for Nonlinear Studies, Los Alamos National Laboratory, Los Alamos, New Mexico 87545, USA\\
\aff{3} Department of Physics, University of Washington, Seattle, WA 98195}
\begin{document}

\maketitle

\begin{abstract}

Recently, in \cite{Zhang2020}, it was found that in rapidly rotating turbulent Rayleigh--B\'enard convection (RBC) in slender cylindrical containers (with diameter-to-height aspect ratio $\Gamma=1/2$) filled with a small-Prandtl-number fluid ($\Pran\approx0.8$), the Large Scale Circulation (LSC) is suppressed and a Boundary Zonal Flow (BZF) develops near the sidewall, characterized by a bimodal PDF of the temperature, cyclonic fluid motion, and anticyclonic drift of the flow pattern (with respect to the rotating frame). This BZF carries a disproportionate amount ($>60\%$) of the total heat transport for $\Pran < 1$ but decreases rather abruptly for larger $\Pran$  to about $35\%$.
In this work, we show that the BZF is robust and appears in rapidly rotating turbulent RBC in containers of different $\Gamma$ and over a broad range of $\Pran$ and $\Ra$. Direct numerical simulations for $0.1 \leq \Pran \leq 12.3$, $10^7 \leq \Ra \leq 5\times10^{9}$, $10^{5} \leq 1/\Ek\leq10^{7}$ and $\Gamma$ = 1/3, 1/2, 3/4, 1 and 2 show that the BZF width $\delta_0$ scales with the Rayleigh number $\Ra$ and Ekman number $\Ek$ as $\delta_0/H \sim \Gamma^{0} \Pr^{\{-1/4, 0\}} \Ra^{1/4} \Ek^{2/3}$ ($\{\Pran<1, \Pran>1\}$) and the drift frequency as $\omega/\Omega \sim \Gamma^{0} \Pr^{-4/3} \Ra \Ek^{5/3}$, where $H$ is the cell height and $\Omega$ the angular rotation rate. The  mode number of the BZF is  1 for $\Gamma \lesssim 1$ and $2 \Gamma$ for $\Gamma$ = \{1,2\} independent of $\Ra$ and $\Pran$. The BZF is quite reminiscent of wall mode states in rotating convection.

\end{abstract}


\section{Introduction}

Turbulent convection driven by buoyancy and subject to background rotation is a phenomenon of great relevance in many physical disciplines, especially in geo- and astrophysics and also in engineering applications. 
In a model system of Rayleigh--B\'enard convection (RBC) \citep{Bodenschatz2000, Ahlers2009, Lohse2010}, 
a fluid is confined in a container where the bottom is heated, the top is cooled, and the vertical walls are adiabatic. The temperature inhomogeneity leads to a fluid density variation which, in the presence of gravity, produces convective fluid motion. When the system rotates with respect to the vertical axis significant modification of the flow occurs owing to the rotational influence including the suppression of the onset of convection \citep{Chandrasekhar1961, Nakagawa1955}, the enhancement or suppression of turbulent heat transport over different ranges of $\Ra$ and $\Pran$ \citep{Rossby1969, Pfotenhauer1987, Zhong1993, Julien1996, Liu1997}, the transformation of thermal plumes into thermal vortices with a rich variety of local structure dynamics \citep{Boubnov1986,Boubnov1990,Hart2002,Vorobieff2002}, and the emergence of robust wall modes before the onset of the bulk mode \citep{Buell1983,Pfotenhauer1987,Zhong1991,Ecke1992,Kuo1993,Herrmann1993,Goldstein1993}. 

The dimensionless control parameters in rotating RBC (RRBC) are the Rayleigh number $\Ra \equiv \alpha g \Delta H^3/(\kappa \nu)$,
the Prandtl number $\Pran\equiv \nu/\kappa$,  the Ekman number $\Ek\equiv\nu/(2\Omega H^2)$,
and the diameter-to-height aspect ratio of the container, $\Gamma\equiv D/H$.
Here $\alpha$ denotes the isobaric thermal expansion coefficient, $\nu$ the kinematic viscosity, $\kappa$ the thermal diffusivity of the fluid,
$g$ the acceleration due to gravity, $\Omega$ the angular rotation rate, $\Delta\equiv T_+-T_-$ the difference between the temperatures at the bottom ($T_+$) and top ($T_-$) plates, 
$H$ the distance between the isothermal plates (the cylinder height), and $D\equiv2R$ the cylinder diameter.
The Rossby number $\Ro \equiv  {\sqrt{\alpha g\Delta H}}\;/\,({2 \Omega H}) = \sqrt{\Ra/\Pran} \Ek$ is another important non-dimensional parameter that provides a measure of the balance between buoyancy and rotation and is independent of dissipation coefficients. 

The important global response parameter in thermal convection is the averaged total heat transport between the bottom and top plates, described by the Nusselt number, 
$\Nu\equiv (\langle u_zT \rangle_{z}-\kappa\partial_z\langle {T}\rangle_{z})/(\kappa\Delta/H)$.
Here, $T$ denotes the temperature, $\bf{u}$ is the velocity field with component $u_z$ in the vertical direction, and $\langle \cdot\rangle_{z}$ denotes the average in time and over a horizontal cross-section at height $z$ from the bottom.

Rotation has various effects on the structure of the convective flow and on the global heat transport in the system.
Rotation inhibits convection and causes an increase of the critical $\Ra_c \sim \Ek^{-4/3}$ at which the quiescent fluid layer becomes unstable throughout the layer \citep{Chandrasekhar1961, Nakagawa1955, Rossby1969, Lucas1983, Zhong1993}. 
In finite containers and sufficiently large rotation rates, however, a different instability occurs at lower $\Ra_w \sim \Ek^{-1}$ in the form of anti-cyclonically drifting wall modes 
\citep{Buell1983, Pfotenhauer1987, Zhong1991, Ecke1992, Ning1993, Zhong1993, Kuo1993, Herrmann1993, Goldstein1993, Goldstein1994, Liu1997, Liu1999, Zhang2009, Favier2020}.  The relative contribution of the wall modes to the total heat transport depends on $\Gamma$ \citep{Rossby1969,Pfotenhauer1987, Zhong1993, Ning1993, Liu1999} with decreasing contribution --- roughly as the perimeter to the area ratio --- with increasing $\Gamma$.   
 
There are several regions of bulk rotating convection where rotation plays an important role, namely a rotation-affected regime and a rotation-dominated regime. In the former where $\Ro \lesssim 1$, heat transport varies as a power law in $\Ra$, i.e., $\Nu = A(\Ek) \Ra^{0.3}$ and can be enhanced or weakly suppressed by rotation relative to the heat transport without rotation \citep{Rossby1969,Zhong1991,Julien1996,Liu1997,King2009,Zhong2009,Liu2009} depending on the range of $\Ra$ and $\Pr$.  In the latter case in which $\Ro \ll 1$, heat transport changes much more rapidly with $\Ra$ in what is known as the geostrophic regime of rotating convection \citep{Sakai1997,Grooms2010,Julien2012,Ecke2014,Stellmach2014,Cheng2020}.

Despite considerable previous work, the spatial distribution of flow and heat transport in confined geometries has not been well studied for high $\Ra$ and low $\Ro$ when one is significantly above the onset of bulk convection but still highly affected by rotation.  Recently, \cite{Zhang2020} demonstrated in direct numerical simulations (DNS) and experiments that a boundary zonal flow (BZF) develops near the vertical wall of a slender cylindrical container ($\Gamma=1/2$) in rapidly rotating turbulent RBC for $\Pran=0.8$ (pressurised gas SF$_6$) and over a broad range of $\Ra$ ($\Ra=10^9$ in DNS and for $10^{11} \lesssim \Ra \lesssim 10^{14}$ in experiments) and $\Ek$ ($10^{-6} \lesssim \Ek \lesssim 10^{-5}$ in the DNS and for $3 \times 10^{-8}  \lesssim \Ek \lesssim 3 \times 10^{-6}$ in experiments). The BZF becomes the dominant mean flow structure in the cell for $\Ro \lesssim 1$, at which the large scale mean circulation (termed the large scale circulation or LSC) vanishes \citep{Vorobieff2002, Kunnen2008, Weiss2011,Weiss2011b}. Further, it contributes a disproportionately large fraction of the total heat transport. Another group \citep{Wit2020} also showed the existence of the BZF and its strong influence on heat transport using DNS for $\Pran=5$ (water) and $\Gamma=1/5$ for $\Ek = 10^{-7}$ in the range $5 \times 10^{10} \lesssim \Ra \lesssim 5 \times 10^{11}$.  Thus, the BZF was observed in different fluids, in cells of different aspect ratios, and over a wide range of parameter values.  Given the strongly enhanced heat transport in the BZF region \citep{Zhang2020, Wit2020}, it is important to explore the BZF in detail.  Here we investigate the robustness of the BZF with respect to $\Pran$ and to $\Gamma$ in the geostrophic regime; we do not address here the transition from the low rotation state to the BZF.

Recently,  \cite{Favier2020} demonstrated for $Ek = 10^{-6}$ through DNS that the linear wall modes of rotating convection \citep{Buell1983, Zhong1991, Ecke1992, Zhong1993, Ning1993, Herrmann1993, Kuo1993, Goldstein1993, Liu1997, Liu1999, Sanchez2005, Horn2017, Aurnou2018} evolve with increasing $\Ra$ and appear to be robust with respect to the emergence of bulk convection even with well developed turbulence.  They suggested that the BZF may be the nonlinear evolution of wall modes, an idea that we address briefly but that requires significantly more analysis and comparison than can be included here.

In the present work, a series of DNS is carried out to study the robustness and the scaling properties of the BZF with respect to Rayleigh number $\Ra$, Ekman number $\Ek$, Prandtl number $\Pran$, and cell aspect ratio $\Gamma$.
We explore the extended scalings of the characteristics of the BZF including the width of the BZF, the drift frequency of the BZF, and the heat transport within the BZF in terms of these non-dimensional parameters. 
We first present our numerical methods, then discuss the results of our calculations, and conclude with our main findings. 


\section{Numerical method}

We present results of direct numerical simulations (DNS) of RRBC in a cylindrical cell obtained using the {\sc goldfish} code \citep{Kooij2018, Shishkina2015} for  $\Ra$ up to $5\times10^{9}$ and $\Ek$ down to $10^{-7}$.
In the DNS, the Oberbeck--Boussinesq approximation is assumed as in \cite{Horn2014}. Centrifugal force effects are neglected
since the Froude number in experiments is typically small, see \cite{Zhong2009, Horn2015}.

The governing equations based on the Oberbeck--Boussinesq approximation are
\begin{eqnarray}
        \nabla\cdot\vec{u}=0,\label{eq:incompressibility}\\
	\partial_t \vec{u} + (\vec{u}\cdot\nabla)\vec{u} &=&- \frac{1}{\rho}\nabla p+ \nu\nabla^2 \vec{u}  - 2\bm{\Omega} \times \vec{u}+\alpha(T-T_0)g\vec{e}_z,\label{eq:momentum}\\
	\partial_t T + (\vec{u}\cdot\nabla) T &=& \kappa\nabla^2 T.\label{eq:energy}
\end{eqnarray}
Here, $\vec{u}=(u_r,u_\phi,u_z)$ is the velocity with radial, azimuthal and vertical coordinates, respectively, $\rho$ is the density, $p$ is the reduced pressure, $\bm{\Omega}=\Omega \vec{e}_z$ is the angular rotation rate vector, $T$ is the temperature with $T_0=(T_++T_-)/2$, and $\vec{e}_z$ is the unit vector in the vertical direction.
Applied boundary conditions are no-slip for the velocity on all surfaces, constant temperature for the top/bottom plates  and adiabatic for the sidewall.
To non-dimensionalise the governing equations, we use ${\Delta=T_+-T_-}$ as the temperature scale, the cylinder height $H$ as the length scale, and the free-fall velocity $\sqrt{\alpha g \Delta H}$ as the velocity scale (the corresponding time scale is $\tau_{ff} = \sqrt{H/\left (\alpha g \Delta \right )}$).

To evaluate the grid requirements for the simulations, we consider the thermal and velocity boundary layers (BLs) near solid boundaries.  The thickness of the BLs near the heated and cooled plates are calculated as
\begin{equation}
\delta_\text{th}= H/(2\Nu).
\label{eq:deltath}
\end{equation}
This is the standard way to define the thermal BL thickness under the assumption of pure conductive heat transport within this layer, cf. \cite{Ahlers2009}.
The viscous BL thicknesses near the plates ($\delta_\text{u}$) and near the sidewall ($\delta^{\text{sw}}$) are defined
as the distances from the corresponding walls to the location where the maxima of, respectively
$\sqrt{<{u_r}^2>_{t, \phi, r}+<{u_\phi}^2>_{t, \phi, r}}(z)$
and
$\sqrt{<{u_\phi}^2>_{t, \phi, z}+<{u_z}^2>_{t, \phi, z}}(r)$
are obtained.
The velocity components are all averaged in time and over the surface parallel to the corresponding wall.
The same criterion was used previously in studies of the sidewall layers in rotating convection, see \cite{Kunnen2011}.

The computational grids are set to be sufficiently fine to resolve the mean Kolmogorov microscales \citep{Shishkina2010} in the bulk and within the BLs (see table \ref{TAB2} in the Appendix).
Grid nodes are clustered near the walls to resolve thermal and velocity BLs resulting in grids that are non-equidistant in both the radial and vertical directions.
As rotation increases, the viscous BL gets thinner \citep{Kunnen2008, Stevens2010, Horn2015} so more points are required near boundaries:  we take at least 7 points within each BL.
The details of all simulated parameters and the corresponding grid resolution are listed in Appendix table \ref{TAB2} along with a benchmark comparison between $\Nu$ data from these simulations and from experimental data in compressed gases with similar $\Pran$ from \cite{Wedi2020}; the agreement is excellent.
To explore the robustness of the BZF with respect to $\Ra$, $\Pran$ and $\Gamma$, we conducted simulations in three groups, i.e., in every group we vary only one parameter while keeping the others fixed. The specific parameter ranges are shown in table \ref{TAB1} (also included in several figures with $\Ra = 10^9$ and $\Pran = 0.8$ are data in the range $0.5 \leq 1/\Ro \leq 5$ from \cite{Zhang2020}; the calculation details for those values are included in the Appendix).
 \begin{table*}
\begin{center}
\begin{tabular}[t]{lcccc}
\toprule
$\Gamma$ & $\Pran$ & $\quad\Ra [10^7]\quad$  & $1/\Ro$ & $\quad \Ek [10^{-6}]\quad$ \\
\hline
1/2             &0.8   			& 5--500 	       	& 10  		&1.3 --13 \\
1/2             &0.1--12.3        	& 10            	&  10 		& 3.2 -- 35 \\
1/3 -- 2        &0.8        		& 10           	&  10 		& 8.9 \\
1/2             &0.8         		& 100          	& 5.6 -- 33.3 	& 0.85 -- 5.1\\                 
 \bottomrule
 \end{tabular}
\caption{Ranges of $\Gamma$, $\Ra$, $\Pran$, $\Ro^{-1}$, and $\Ek$. For details, see Appendix.}
\label{TAB1}
\end{center}
\end{table*}



\section{Results}

\subsection{Boundary Zonal Flow structure}

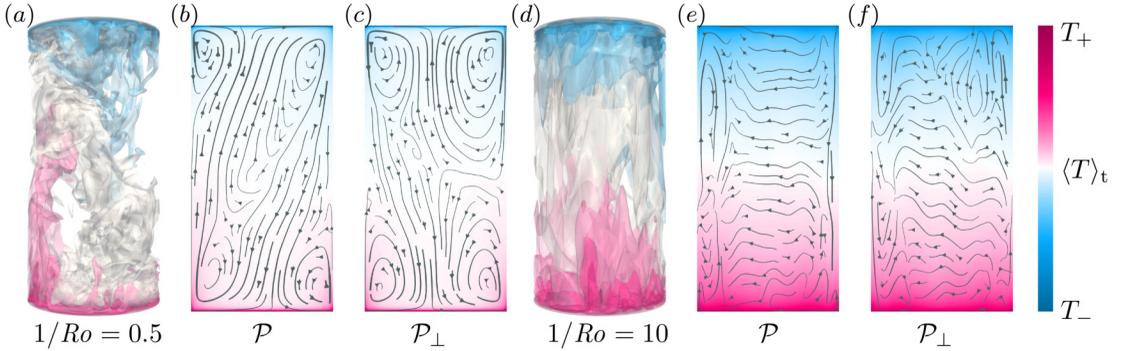
\begin{figure*}
\unitlength1truecm
\begin{picture}(15,4.5)
\put(-0.5, 0){\includegraphics[width=15cm]{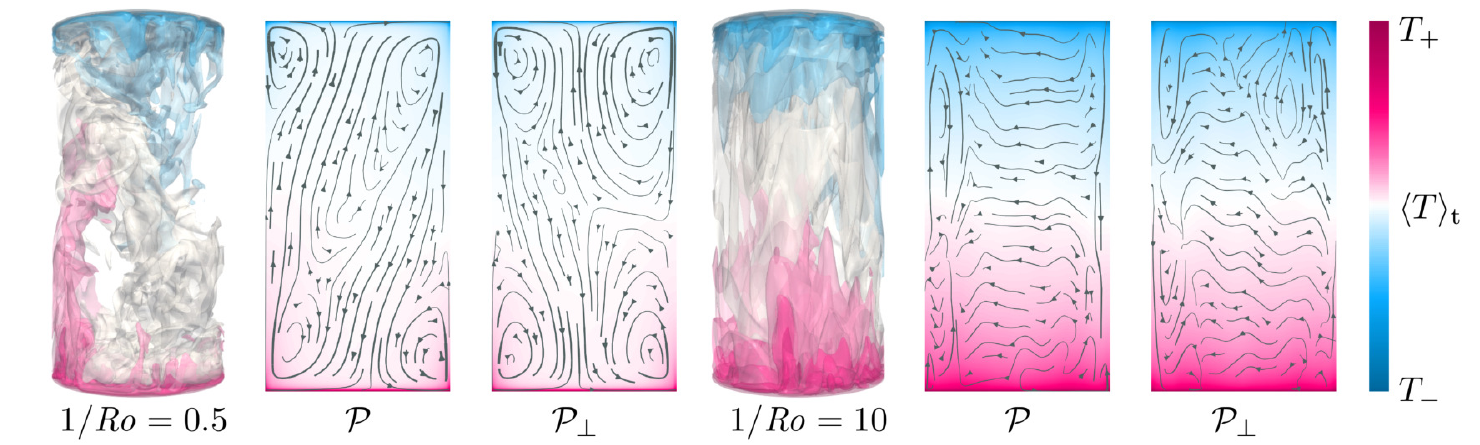}}
\put(-1.0,-0.15){
\put(0.7,4.5){$(a)$}
\put(2.9,4.5){$(b)$}
\put(5.2,4.5){$(c)$}
\put(7.4,4.5){$(d)$}
\put(9.6,4.5){$(e)$}
\put(11.85,4.5){$(f)$}
}
\end{picture}
\caption{
Isosurfaces of  instantaneous temperature $T$ $(a)$ and time-averaged flow fields $(b,\,c)$,
visualised by streamlines (arrows) and temperature (colours),
for  $\Ra=10^9$ and  $1/\Ro=10$,
in vertical orthogonal planes $\mathcal{P}$ $(b,\,e)$ and $\mathcal{P}_\perp$ $(c,\,f)$.
In the case of weak rotation $(a,\,b,\,c)$, $\mathcal{P}$ is the plane of the large-scale circulation $(b)$. 
Averaging in $(b,\,c)$ is conducted over 1000 free-fall time units. 
For strong rotation $(e,\,f)$,  mean radial and axial velocity magnitudes are approximately tenfold smaller than those for weak rotation  $(b,\,c)$.
}
\label{PIC1}
\end{figure*}

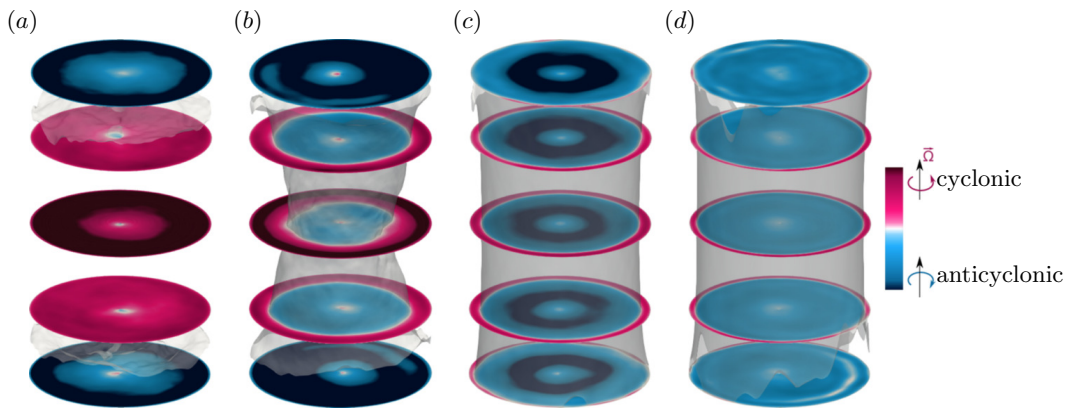
\begin{figure*}
\unitlength1truecm
\begin{picture}(12,5.2)
\put(0.5, 0){\includegraphics[width=12cm]{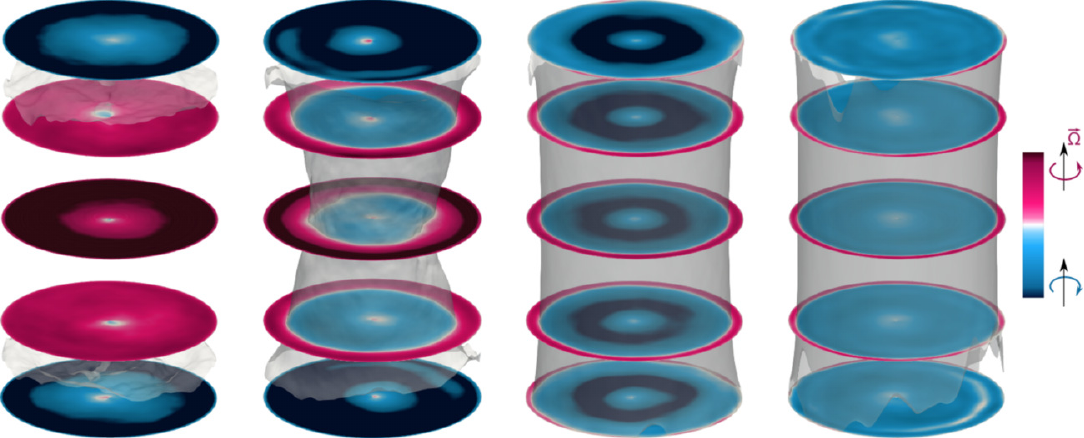}}
\put(-0.3,0){
\put(0.5,5){$(a)$}
\put(3.5,5){$(b)$}
\put(6.4,5){$(c)$}
\put(9.2,5){$(d)$}
}
\put(12.5, 2.92){cyclonic}
\put(12.5, 1.65){anticyclonic}
\end{picture}
\caption{Time-averaged fields $\langle u_{\phi}\rangle_t$ for $\Pran=0.8$, $\Gamma=1/2$, $\Ra=10^9$ and 
$(a)$ $1/\Ro=0.5$, $(b)$ $1/\Ro=2$, $(c)$ $1/\Ro=10$, $(d)$ $1/\Ro=20$. 
}
\label{PIC2}
\end{figure*}

Our goal here is to explore the robustness of the BZF with respect to variations of control parameters.  We follow closely the approach and characterization presented in \cite{Zhang2020} but focus on the geostrophic regime where the BZF is well developed. After presenting our main results, we consider the BZF with respect to wall mode structures.  We begin with the influence of rotation on the overall temperature and velocity fields in the cell.
In figure \ref{PIC1}, for particular cases of $1/\Ro=0.5$ (weak rotation) and $1/\Ro=10$ (fast rotation),
3D instantaneous temperature distributions (figure \ref{PIC1}a, d) and 2D vertical cross-sections (figure \ref{PIC1}b,c,e,f) of the time-averaged flow fields are shown.
The 2D views are taken in a plane $\mathcal{P}$ (figure \ref{PIC1} b, e), which in the case of a weak rotation is the LSC plane,
and additionally in a plane $\mathcal{P}_\perp$ which is perpendicular to $\mathcal{P}$ (figure \ref{PIC1} c, f).
For slow rotation, a LSC spanning the entire cell with 2 secondary corner rolls are observed in $\mathcal{P}$
whereas a 4-roll structure is seen in $\mathcal{P}_\perp$, typical of classical RBC at large $\Ra$ and for $\Gamma \sim 1$ (e.g., see \cite{Shishkina2014} and \cite{Zwirner2020}).
Near the plates, the LSC and the secondary corner flows move the fluid towards the sidewall (figure \ref{PIC1}b) so the Coriolis acceleration ($-2\Omega{\bf e}_z\times{\bf u}$) induces anticyclonic fluid motion close to the plates. 
In the central part of the cell, at $z=H/2$, the radial component of the mean velocity, $\langle u_r\rangle_t$, 
always points towards the cell center (figure \ref{PIC1}a, b).
Therefore, Coriolis acceleration results in cyclonic fluid motion in the central part of the cell as is also observed in the time-averaged azimuthal velocity field
$u_{\phi}$ in figure \ref{PIC2}a.
Cases at higher rotation rates are shown in figures \ref{PIC1}d-f and \ref{PIC2} (see also \cite{Kunnen2011}).
For both small and large rotation rates, the presence of viscous BLs near the plates implies anticyclonic motion of the fluid there.
For strong rotation, the subject of this paper, with high and constant angular velocity $\Omega$, 
the fluid velocity becomes more uniform along ${\bf e}_z$ owing to the Taylor--Proudman constraint with larger components of lateral velocity compared to the vertical component as in figures  \ref{PIC1}e, f.
Thus, anticyclonic fluid motion is present not only in the vicinity of the plates,
but involves more and more fluid volume with increasing $\Ro^{-1}$.
With increasing rotation rate, anticyclonic motion grows from the plates toward the cell center whereas cyclonic motion at $z=H/2$ remains near the sidewall and becomes increasingly more localized there (figures \ref{PIC2}c, d).

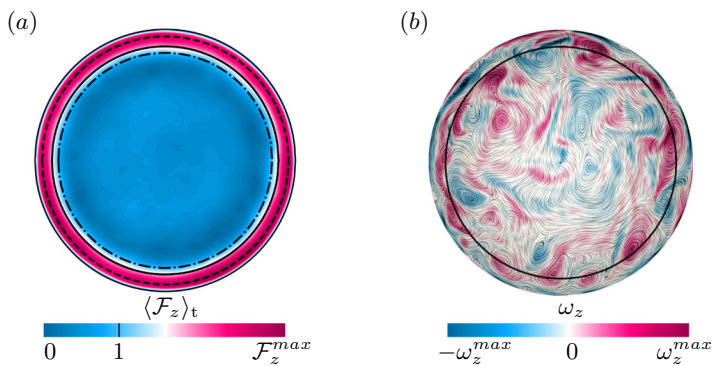
\begin{figure*}
\unitlength1truecm
\begin{picture}(12,4.5)
\put(2, 0){\includegraphics[width=9cm]{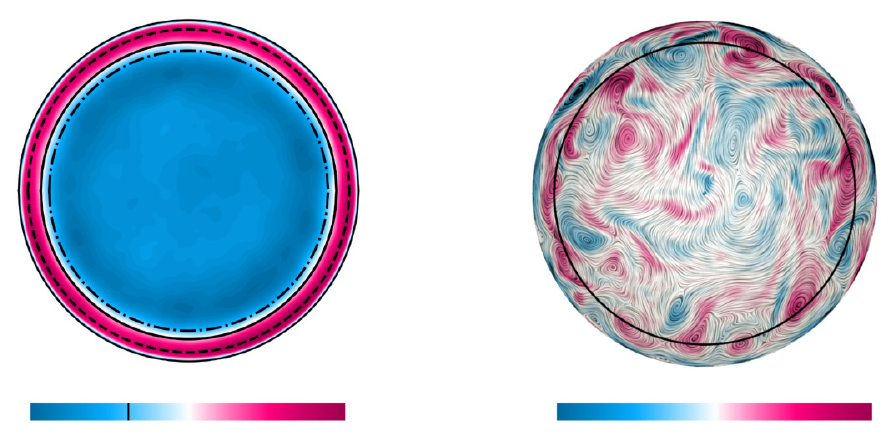}}
\put(1.8, 4.3){$(a)$}
\put(7, 4.3){$(b)$}
\put(3.6,0.55){$\avg{\mathcal{F}_z}{t}$}
\put(9.1,0.55){$\omega_z$}
\put(7.5, -0.05){$-\omega_z^{max}$}
\put(9.2, -0.05){$0$}
\put(10.4, -0.05){$\omega_z^{max}$}

\put(2.3, -0.05){$0$}
\put(3.21, -0.05){$1$}
\put(5.05, -0.05){$\mathcal{F}_z^{max}$}

\end{picture}
\caption{For $\Ra=10^9$, $1/\Ro=10$, $\Pran=0.8$, $\Gamma=1/2$, and $z = H/2$:
$(a,\,b)$ Horizontal cross-sections of $(a)$ time-averaged vertical heat flux $\avg{\mathcal{F}_z}{t}$
and $(b)$ instantaneous  vertical component of vorticity $\omega_z$ (negative values correspond to anticyclonic fluid motion),
together with two-dimensional streamlines. The solid line indicates the radial position $r_0$ that defines the BZF by the condition $\avg{u_\phi \left (r_0, z=H/2 \right )}{t}=0$. In $(a)$, the dash-dotted line (inner circle) and the dashed line (outer circle) are, respectively, the radial locations of $\avg{\mathcal{F}_z}{t}=1$ (global averaged heat flux) and $u_\phi^{\max}$, the maximum of  the time-averaged azimuthal velocity.
}
\label{PIC3}
\end{figure*}

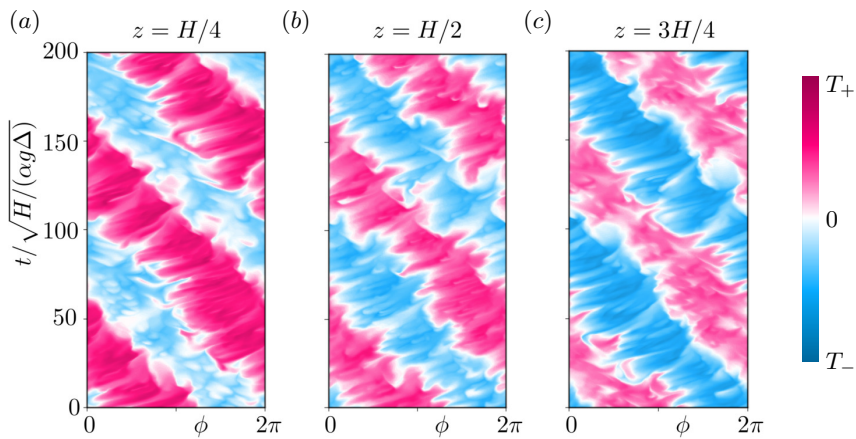
\begin{figure*}
\unitlength1truecm
\begin{picture}(12, 5.5)
\put(1, 0){\includegraphics[width=10cm]{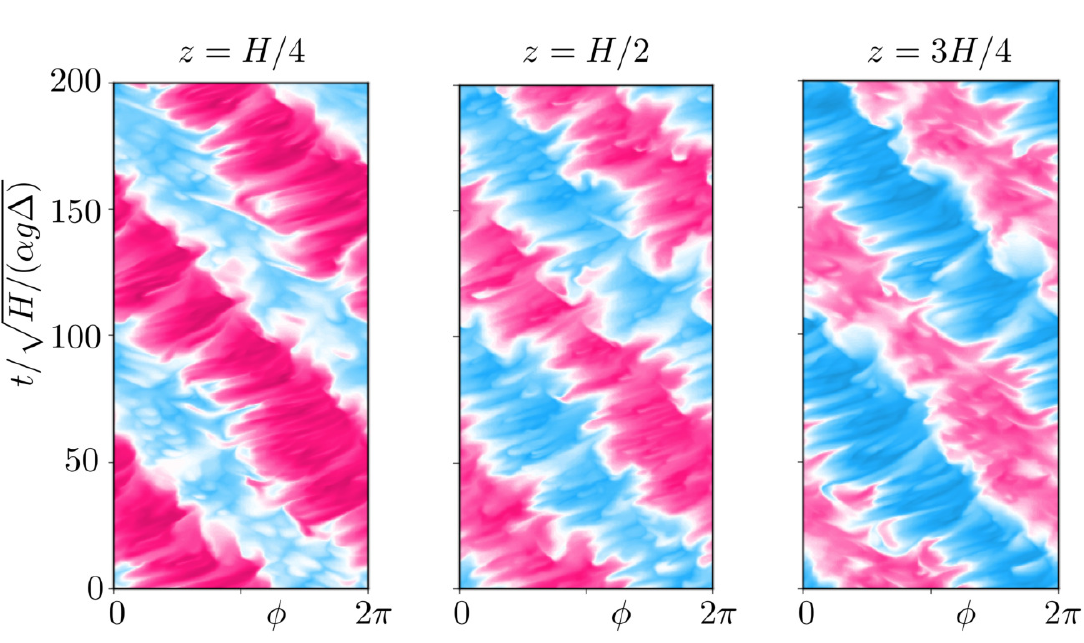}}
\put(1, 5.5){$(a)$}
\put(4.6, 5.5){$(b)$}
\put(7.8, 5.5){$(c)$}
\put(11.5,1.1){\includegraphics[height=2mm, width=3.75cm, angle = 90 ]{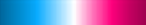}}
\put(11.8,1.0){$T_-$}
\put(11.8,2.88){$0$}
\put(11.8,4.65){$T_+$}
\end{picture}
\caption{
For $\Pran=0.8$, $\Gamma=1/2$, $\Ra=10^9$, $1/\Ro=20$, and $r=R$:
time evolution of temperature distribution (space-time plot of temperature) at height $(a)$ $z=H/4$, $(b)$ $z=H/2$, $(c)$ $z=3H/4$.
}
\label{PIC4}
\end{figure*}

\begin{figure*}
\unitlength1truecm
\begin{picture}(12, 9.5)
\put(1, 4){\includegraphics[width=12cm]{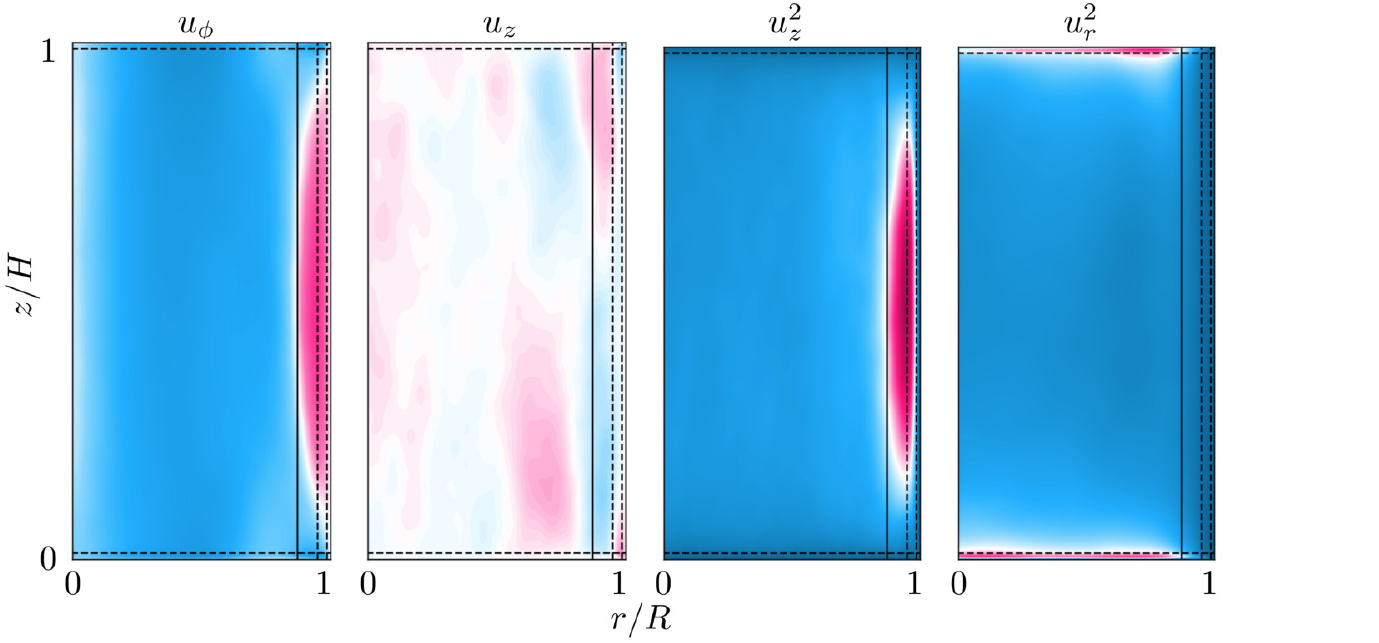}}
\put(0,4.1){
\put(1, 5.3){$(a)$}
\put(3.8, 5.3){$(b)$}
\put(6.4, 5.3){$(c)$}
\put(9, 5.3){$(d)$}
}

\put(5.2,0){
\put(2.7, 0.35){\includegraphics[height=2.5mm, width=2.6cm, angle = 0]{figs/fig-tscale.pdf}}
\put(2.6, 0.08){0}
\put(5, 0.08){max}
}

\put(2.7, 0.35){\includegraphics[height=2.5mm, width=2.6cm, angle = 0]{figs/fig-tscale.pdf}}
\put(2.45, 0.08){min}
\put(3.95, 0.08){0}
\put(5, 0.08){max}

\put(1, 0.5){\includegraphics[width=12cm]{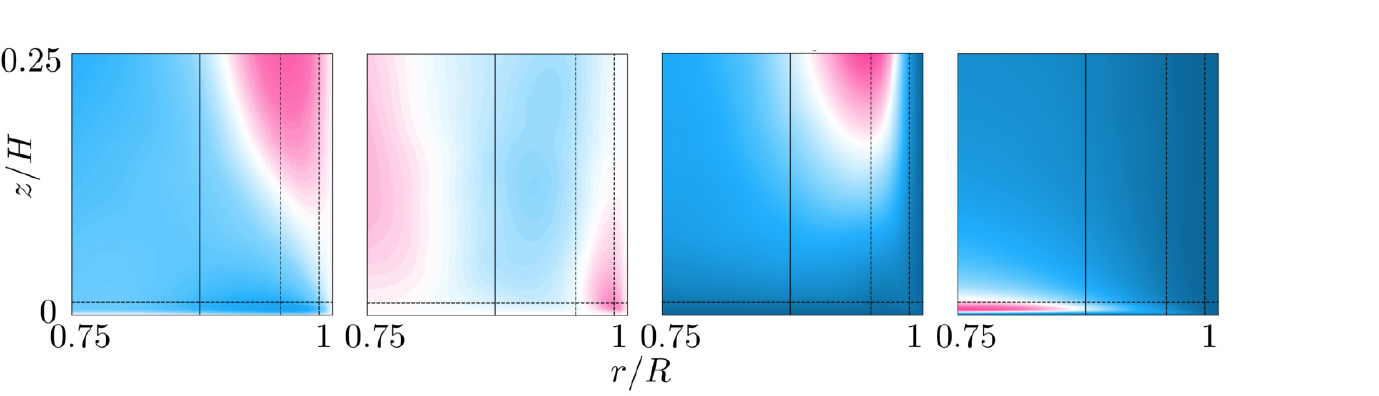}}
\put(0,-1.5){
\put(1, 5.3){$(e)$}
\put(3.8, 5.3){$(f)$}
\put(6.4, 5.3){$(g)$}
\put(9, 5.3){$(h)$}
}
\end{picture}
\caption{
Time-averaged flow fields in a vertical plane, for $\Ra=10^9$, $1/\Ro=10$, $\Pran=0.8$, $\Gamma=1/2$.
Range of variables are, respectively: $(a, b, e, f)$ from -0.17 to 0.17; 
$(c, d, g, h)$ from 0 to 0.0289.}
\label{PIC5}
\end{figure*}

\begin{figure*}
\unitlength1truecm
\begin{picture}(13, 9.5)
\put(0,0.5){
\put(2, 4.0){\includegraphics[width=10cm]{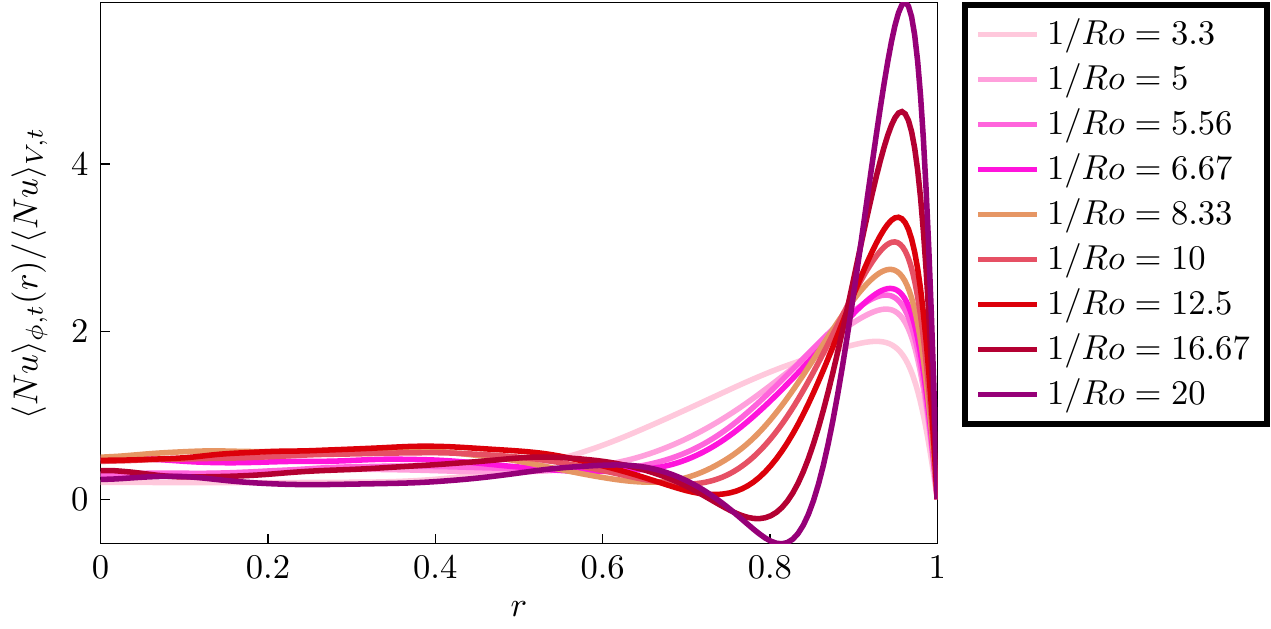}}
\put(3, 5.6){\includegraphics[width=5cm]{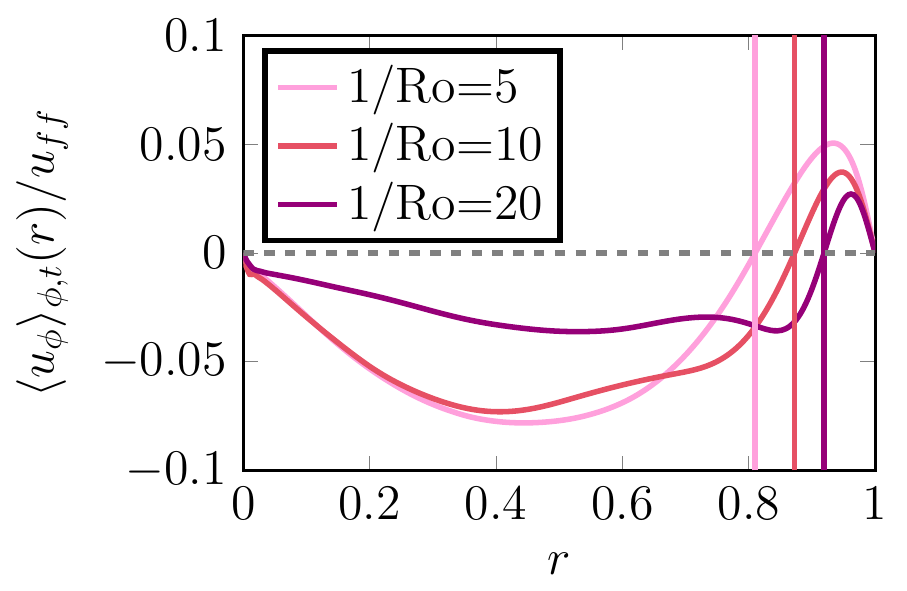}}
}
\put(-0.5, -0.25){\includegraphics[width=5.1cm]{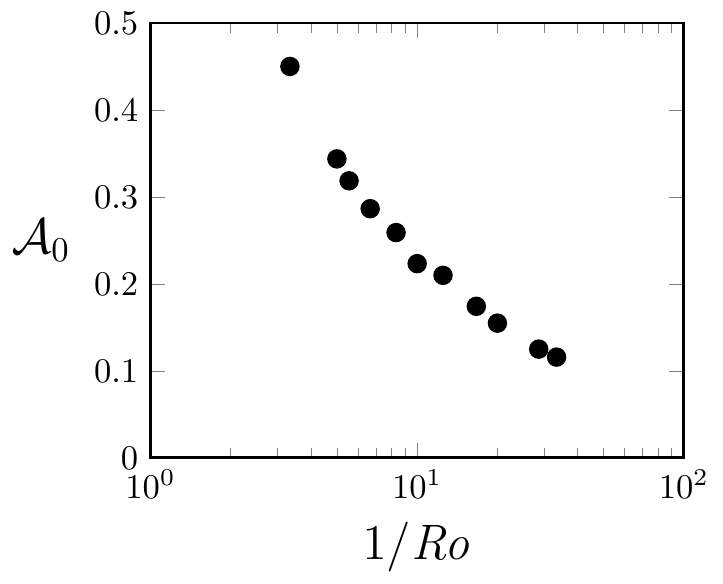}}
\put(4.3, -0.25){\includegraphics[width=4.9cm]{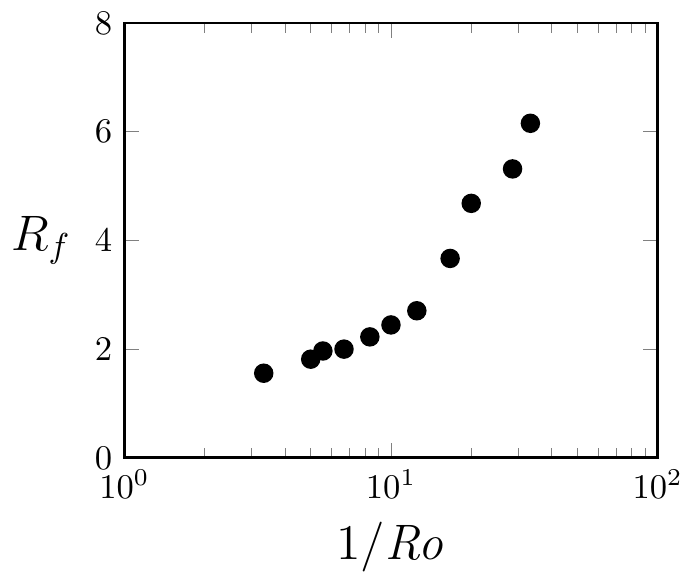}}
\put(8.95, -0.25){\includegraphics[width=5.1cm]{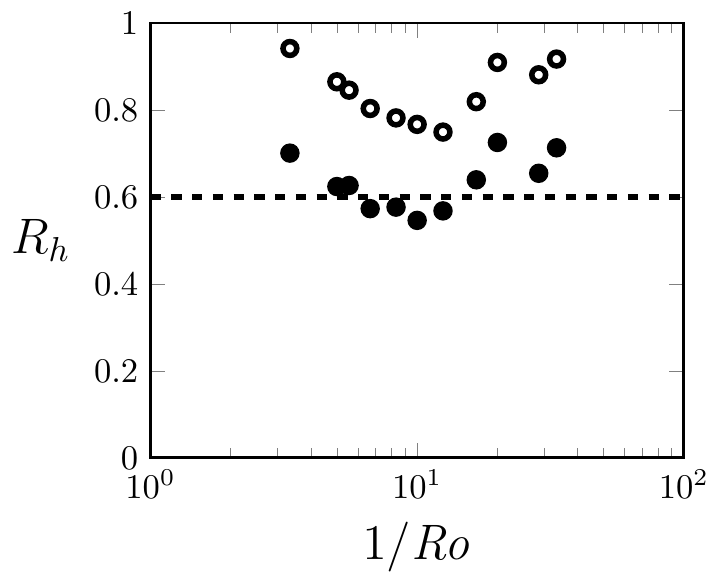}}
\put(11.5, 0.7){\includegraphics[width=2.2cm]{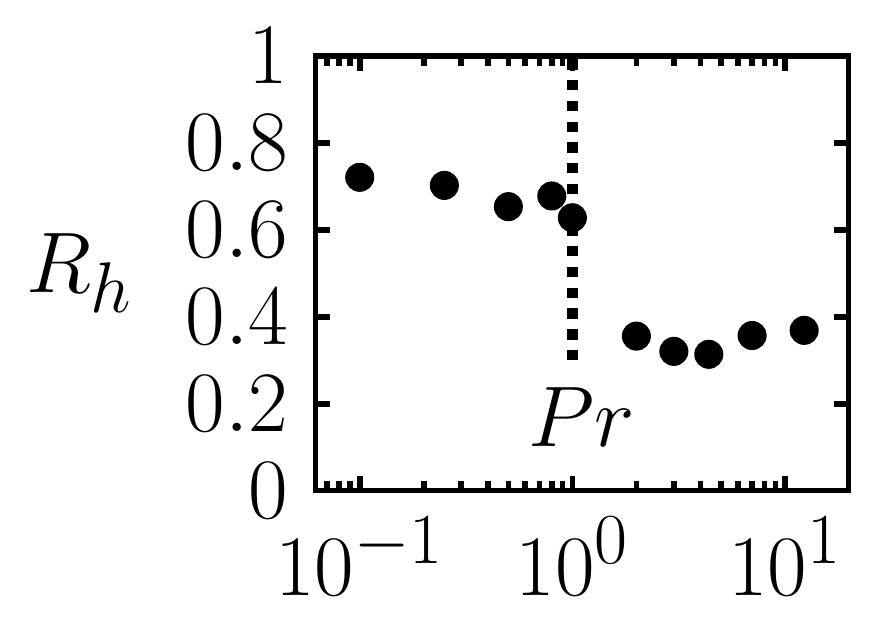}}
\put(2, 9.5){$(a)$}
\put(-0.3, 4.0){$(b)$}
\put(4.5, 4.0){$(c)$}
\put(9.4, 4.0){$(d)$}
\end{picture}
\caption{
$(a)$ Radial profiles of normalised time- and $\phi$- averaged heat flux $\langle \Nu \rangle_{\phi,t}(r)/\langle \Nu \rangle_{V,t}$ at $z=H/2$, for different rotation rates.
Inset shows the radial profiles of time- and $\phi$- averaged $u_{\phi}$, where solid lines pass $\langle u_{\phi} \rangle_{\phi,t}=0$ (radial location corresponds to $r_0$).
$(b)$ Ratio of BZF area to the total area at $z=H/2$, $\mathcal{A}_0 = (R^2-r_0^2)/R^2$;
$(c)$ Ratio of mean vertical heat flux inside BZF to mean global heat flux, $R_f$, equation (\ref{eq:Rf});.
$(d)$ Ratio of heat transported inside BZF (solid circles) or in an extended zone $R-2\delta_0 < r < R$ (open circles) to total transported heat, $R_h$ ($R_h^*$), equation (\ref{eq:Rh}).
Everywhere $\Ra=10^9$, $\Pran=0.8$, $\Gamma=1/2$.
}
\label{PIC6}
\end{figure*}

As introduced in \cite{Zhang2020}, the BZF in rapidly rotating turbulent convection is characterised by an anticyclonic bulk flow, cyclonic vortices clustering near the sidewall, and anticyclonic drift of thermal plumes (see figures \ref{PIC3}a,b and figure \ref{PIC4}).
These structures are associated with the bimodal temperature PDFs obtained in the measurements and DNS near the sidewall \citep{Zhang2020, Wedi2020}.
The radial location $r_0$ where the mean fluid motion at $z/H=1/2$ changes from anticyclonic to cyclonic as indicated by the solid line in figure \ref{PIC3} (see also inset of figure \ref{PIC6}a below) is used to describe the width of the BZF $\delta_0 = R - r_0$. 
As one might expect, vertical coherence of the BZF is enhanced by strong rotation. 
In figure \ref{PIC4}, time-angle plots of the temperature at 3 different heights show that the drift frequency $\omega = 2 \pi R(d\phi (r_{u_\phi^{max}})/ dt)/(2 \pi R/m) =  m d\phi (r_{u_\phi^{max}})/ dt$ is quite constant along $z$ without significant phase differences, i.e., the BZF maintains good vertical coherence.
Here, the mode number $m$ equals 1 and $d\phi (r_{u_\phi^{max}})/ dt$ denotes the angular velocity of the temperature drift at $r=r_{u_\phi^{max}}$, where the maximum of the time-averaged azimuthal velocity is obtained.
In the lower half of the cell, for $z=H/4$, warm plumes dominate so the warm regions (pink stripes) are wider, whereas in the upper half of the cell, for $z=3H/4$, cold plumes dominate resulting in wider cooler regions (blue stripes). 
Similarly, figures \ref{PIC2}c,d and \ref{PIC5}a,d show that the zonal flow develops away from the top/bottom plates and extends vertically throughout the bulk.  
Figure \ref{PIC5} illustrates that owing to the drift, time-averaged fields in the vertical plane average to zero and do not capture important features of the flow motion, in particular, the $u_z$-field. The averaged $u^2_z$, however, does retain important information about the locations of the Stewartson `1/3' and `1/4' layers (dashed lines) and the BZF (solid line).

\subsection{Contribution to heat transport}

An important and unexpected property of the BZF in rotating RBC is its disproportionately large contribution to the heat transport in the system. Figures \ref{PIC3}a and~\ref{PIC6} show that the averaged heat flux inside the BZF is much stronger than in the region outside the BZF. To be clear about the averaging we define
\begin{eqnarray}
\mathcal{F}_i(r, \phi, z) &\equiv& (u_iT-\kappa\partial_iT)/(\kappa\Delta /H),\, i=r, \phi, z, \\
\langle \Nu (r, t) \rangle_\phi &\equiv& (2 \pi)^{-1} \int_0^{2 \pi} \mathcal{F}_z(r, \phi, z=H/2) d\phi, \\
\langle \Nu (t) \rangle_{V} &\equiv&(\pi R^2 H)^{-1} \int_0^{2\pi} \int_0^R \int_0^H \mathcal{F}_z (r, \phi,z) r dr d\phi dz, \\
\langle \Nu (t) \rangle_{BZF} &\equiv& (\pi (R^2-r_0^2))^{-1}\int_0^{2\pi} \int_{r_0}^R \mathcal{F}_z(r,\phi, z=H/2) r dr d\phi, \\[1pt]
R_f &\equiv& \langle \Nu \rangle_{\text{BZF},t}/\langle \Nu \rangle_{V,t}, \label{eq:Rf}\\[8pt]
R_h &\equiv& (\langle \Nu \rangle_{\text{BZF},t} \cdot \pi (R^2-r_0^2))/(\langle \Nu \rangle_{V,t} \cdot \pi R^2) \nonumber  \\[2pt]
&=& \frac{R^2-r_0^2}{R^2} \langle \Nu \rangle_{\text{BZF},t}/\langle \Nu \rangle_{V,t}, \label{eq:Rh}
\end{eqnarray}
where $r_0=R-\delta_0$.
The quantity $R_f$ is the ratio of the mean vertical heat flux within the BZF to the vertical heat flux averaged over the whole cell.
The quantity $R_h$ reflects the portion of the heat transported through the BZF compared to the total transported heat.
Especially, in figure \ref{PIC6}a, the time- and $\phi$-averaged radial profile at the mid-height for $\Ra = 10^9$, $\Pran = 0.8$, and $\Gamma = 1/2$ shows a significant peak of heat transport inside the BZF, and the peak amplitude increases dramatically as rotation becomes stronger. Thus, although the width of the BZF shrinks with increasing rotation, thereby reducing the effective area of the BZF with respect to the whole domain, as shown in figure \ref{PIC6}b, the increasing magnitude of the peak makes the heat transport carried by the BZF quite significant. Note that the annular BZF region of width  $\delta_0$ is smaller than the positive contribution to the heat transport as shown in the inset of figure \ref{PIC6}a. Figure \ref{PIC6}c reveals that the enhancement of the local heat transfer within the BZF increases more rapidly when rotation is very strong ($1/\Ro \gtrsim 10$). As a result of these properties, the heat transport carried by the BZF for these parameter values is always more than 60\% of the total heat transport at fast rotation (see figure \ref{PIC6}d). Note, however, the effect  of the BZF on the heat transport extends over a wider range $r < r_0$; over some range $Nu$ is actually negative (see figure \ref{PIC6}a) implying an anti-correlation of vertical velocity and buoyancy, i.e., warm fluid going down or cooler fluid moving up.
If we modify the annular averaging to take into account the decreased $Nu$ region as well as the inner structure of the BZF, i.e., we average over the extended region $R - 2\delta_0 \leq r \leq R$, we get the ratio $R_h^*$ which is also shown in figure \ref{PIC6}d (open symbols) where one sees an even larger fractional contribution.

We also consider the dependence of the heat transport ratio $R_h$ as a function of $\Pran$, see inset of figure \ref{PIC6}d. Interestingly, for $\Pran <1$ we find $0.6 < R_h < 0.7$, whereas for $\Pran > 1$ we have $0.3 < R_h < 0.4$ with a quite sharp transition for $\Pran \approx 1$. The origin of this rather sharp change emphasizes the important role $\Pran$  plays, perhaps through the competition between thermal and viscous BLs. Finally, comparing our computation of the total $Nu$ with increasing rotation with that of \cite{Wedi2020} (see Appendix figure \ref{PIC13}), we conclude, given the close agreement, that the contribution of the BZF affects both measures of $Nu$ substantially and needs to be taken into account when considering the scaling of geostrophic heat transport in experiments and also in DNS with no-slip sidewall boundary conditions (see also \cite{Wit2020}).


\begin{figure*}
\unitlength1truecm
\begin{picture}(12, 5.5)
\put(1, 0){\includegraphics[width=10cm]{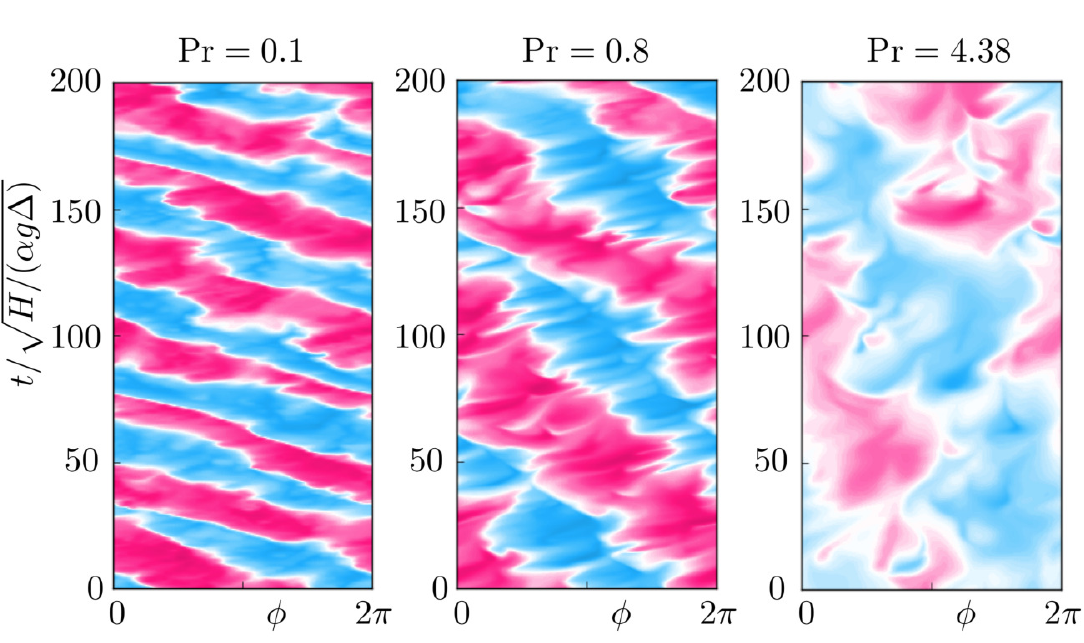}}
\put(1, 5.5){$(a)$}
\put(4.5, 5.5){$(b)$}
\put(7.5, 5.5){$(c)$}
\put(11.5,1.1){\includegraphics[height=2mm, width=3.75cm, angle = 90 ]{figs/fig-tscale.pdf}}
\put(11.8,1.0){$T_-$}
\put(11.8,2.88){$0$}
\put(11.8,4.65){$T_+$}
\end{picture}
\caption{
Space-time plots of temperature $T$ at sidewall, $r=R$, and half-height, $z=H/2$, for $\Ra = 10^8$, $1/\Ro = 10$, $\Gamma = 1/2$, $(a)$ $\Pran=0.1$, $(b)$ $\Pran=0.8$, $(c)$ $\Pran=4.38$.
}
\label{PIC7}
\end{figure*}

\begin{figure*}
\unitlength1truecm
\begin{picture}(12, 5.5)
\put(1, 0){\includegraphics[width=10cm]{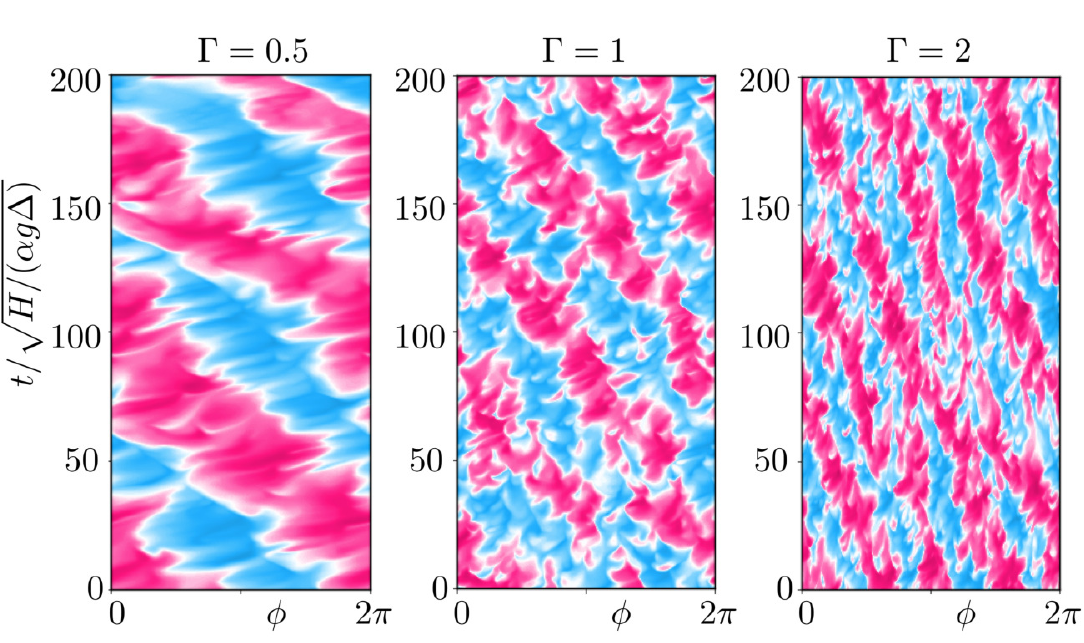}}
\put(1, 5.5){$(a)$}
\put(4.7, 5.5){$(b)$}
\put(7.9, 5.5){$(c)$}
\put(11.5,1.1){\includegraphics[height=2mm, width=3.75cm, angle = 90 ]{figs/fig-tscale.pdf}}
\put(11.8,1.0){$T_-$}
\put(11.8,2.88){$0$}
\put(11.8,4.65){$T_+$}
\end{picture}
\caption{
Space-time plots of temperature $T$ at sidewall, $r=R$, and half-height, $z=H/2$, for $\Ra=10^8$, $1/\Ro=10$, $\Pran=0.8$, $(a)$ $\Gamma=1/2$, $(b)$ $\Gamma=1$, $(c)$ $\Gamma=2$.}
\label{PIC8}
\end{figure*}

\subsection{Dependence on $\Ra$, $\Pran$ and $\Gamma$}

We first discuss the qualitative robustness of the BZF with respect to $\Ra$, $\Pran$ and $\Gamma$ before we consider its quantitative spatial and temporal properties. We demonstrate the character of the BZF with respect to variations of $\Pran$ and $\Gamma$ by considering time-angle plots of temperature $T$ at $z=H/2$ and $r=R$. Figure \ref{PIC7}a shows that the BZF exists in the flows at different $\Pran= 0.1, 0.8, 4.38$ (also for $\Pran= 0.25, 0.5, 2, 3, 7, 12.3$, not shown), i.e., from small to large $\Pran$. Although there are some quantitative differences among the three cases, they all qualitatively demonstrate the existence of the BZF for more than two decades of $\Pran$.

The qualitative dependence of the BZF on the aspect ratio $\Gamma$ is shown in figure \ref{PIC8} for three different aspect ratios: $\Gamma = 1/2, 1, 2$. The BZF is present in all three cases, has the same scaling of BZF width when scaled by $H$, i.e., $\delta_0/H$ is independent of $\Gamma$ (see figure \ref{PIC9}d inset), and has a drift period (in units of free fall time $\tau_\text{ff} = \sqrt{H/\left (\alpha g \Delta\right )} = \tau_\kappa \Pran^{-1/2} \Ra^{-1/2}$ where $\tau_\kappa = H^2/\kappa$ is the thermal diffusion time) of about 70.  The quantitative scaling of the drift frequency is analyzed later, and the data are tabulated in the Appendix (see table \ref{TAB2}). The wavelength $\lambda$ of the traveling BZF mode is independent of $\Gamma$ for these three values in a straightforward way, as seen in figure \ref{PIC8}, namely $\lambda/H = \pi/2$ so that the number of wavelengths around the circumference is $m=2 \Gamma$ and the wavenumber is $k=2\pi/\lambda= 4/H$. We note, however, that this relationship is for a limited number of values of $\Gamma$ and control parameters $\Ra$ and $\Ro$.  Thus, we make no strong claims to its generality.  Indeed, there is already evidence from \cite{Wit2020} that for $\Gamma = 1/5$ one gets $m=1 \neq 2\Gamma$, and we made additional measurements with $\Gamma =$ 1/3 and 3/4 that also yield $m=1$.   We conjecture that owing to periodic azimuthal symmetry $m$ will take on only integer values, similar to the situation for wall mode states  \citep{Ecke1992,Zhong1993,Goldstein1993,Ning1993,Liu1999} in cylindrical convection cells. Because of this periodic constraint, one cannot have $m<1$ so small aspect ratios with $\Gamma \lesssim 1$ have $m=1$. We also note that the mode number dependence on $\Gamma$ of the BZF is similar to that of the $\Gamma$ dependence of linear wall state mode number, i.e., $m \approx 3 \Gamma$ \citep{Goldstein1993,Kuo1993,Herrmann1993,Ning1993,Liu1999,Zhang2009}. Given that our states have values of $\Ra$ that are 10 to 100 times greater than the linear wall mode onset $\Ra_w$, this difference is not unreasonable and the correspondence is very suggestive. 
In particular, a range of mode numbers are stable near onset \citep{Zhong1993,Ning1993,Liu1999} owing to the azimuthally periodic boundary conditions. Significantly above onset there seems to be a selection towards lower mode numbers: e.g., \cite{Zhong1993} figures 3 and 8 with  $\Gamma = 2$ show stable wall modes with $m = 4, 5, 6, 7$ near onset but only the $m=4, 5$ modes persist for higher $\Ra$  which yields $m = 2\Gamma$ and $m = 2.5 \Gamma$, respectively, consistent with our results for the BZF (see also \cite{Favier2020}).

\subsection{Spatial and temporal scales}

\begin{figure*}
\unitlength1truecm
\begin{picture}(12, 15)
\put(0,4.2){
\put(0, 6){\includegraphics[width=6.77cm]{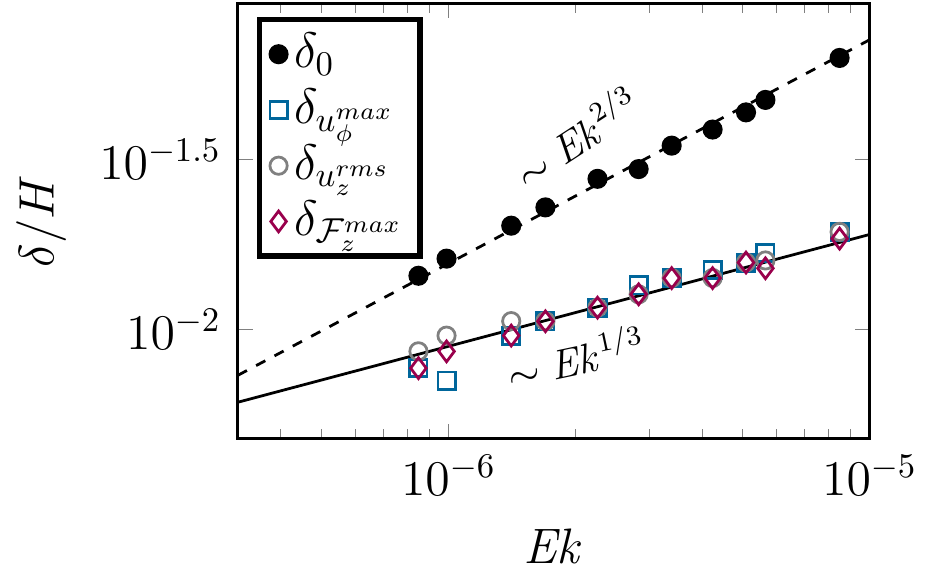}}
\put(6.6, 6){\includegraphics[width=6.7cm]{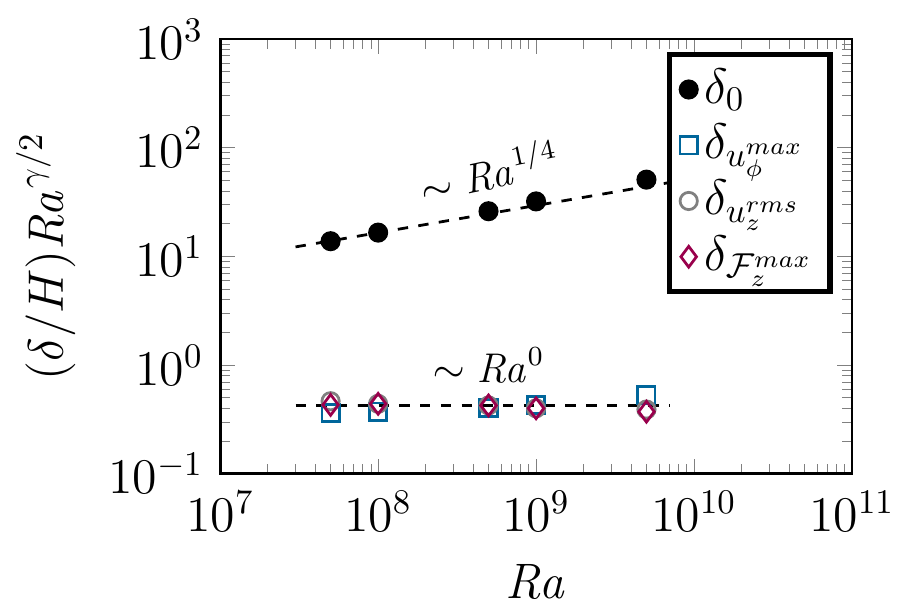}}
\put(0, 1){\includegraphics[width=6.7cm]{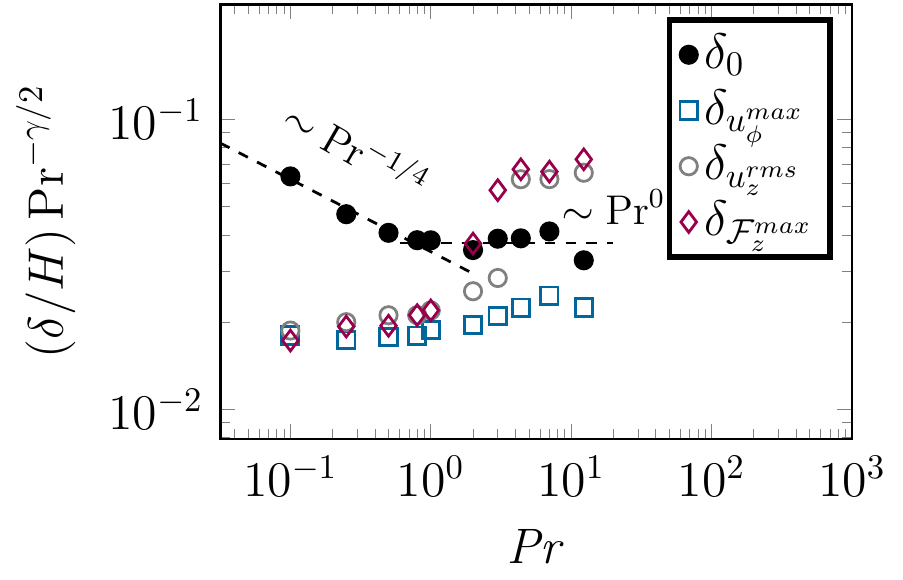}}
\put(6.6, 1){\includegraphics[width=6.32cm]{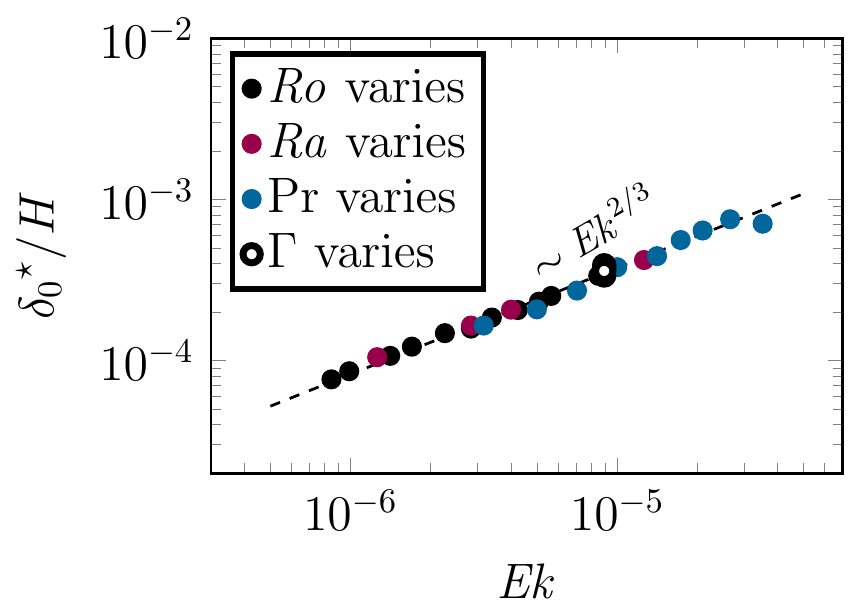}}
\put(10.7, 2.05){\includegraphics[width=2.1cm]{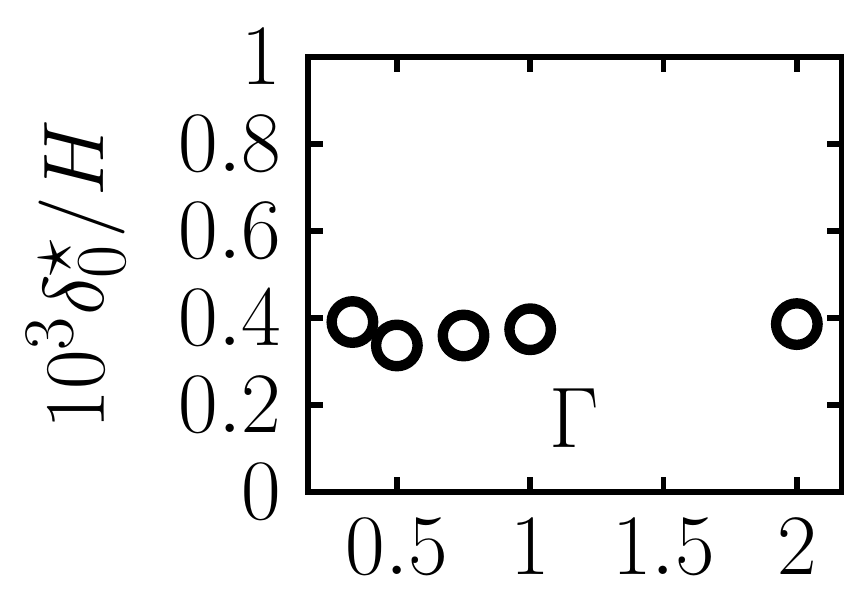}}
\put(0, 10.5){$(a)$}
\put(6.8, 10.5){$(b)$}
\put(0, 5.5){$(c)$}
\put(6.8, 5.5){$(d)$}
}
\put(3,0){\includegraphics[width=6.5cm]{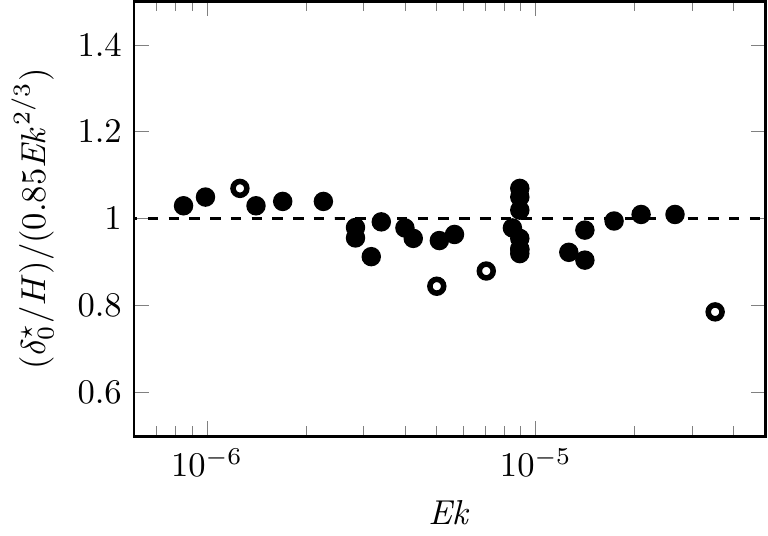}}
\put(2.5, 4.5){$(e)$}
\end{picture}
\caption{
$(a)$ Scaling with $\Ek$ of characteristic widths $\delta/H \sim \Ek^{\gamma}$ (for $\delta_0$: distance from vertical wall to location where $\avg{u_{\phi}}{t}=0$), for $\Ra=10^9$, $\Pran=0.8$, $\Gamma=1/2$. For $\delta_0/H$, $\gamma \sim 2/3$ whereas for other $\delta/H$, $\gamma = 1/3$. 
$(b)$ Scaling with $\Ra$ of compensated width $\Ra^{\gamma/2} \delta_0/H$ for fixed $1/\Ro=10$ and $\Pran =0.8$.
$(c)$ Scaling with $\Pran$ of compensated width $\Pran^{-\gamma/2} \delta_0/H$ for $\Ra=10^8$ and $1/\Ro = 10$.
$(d)$ Scaling with $\Ek$ of normalized BZF width $\delta_0^\star/H = \Ra^{-1/4} \Pran^{1/4} \delta_0/H$ (for $\Pran < 1$) and
$\delta_0^\star/H = \Ra^{-1/4} \Pran^{0} \delta_0/H$ (for $\Pran > 1$); for compactness, we write the two scalings with $\Pran$ as $\Pran^{\{-1/4,\ 0\}}$. The inset shows $\delta_0^\star/H$ vs $\Gamma$.
$(e)$ Compensated plot of BZF width $(\delta_0/H)/(0.85 \Pr^{\{-1/4,\ 0\}}\Ra^{1/4} \Ek^{2/3})$ vs. $\Ek$
(all data from table \ref{TAB2} are shown; open symbols are cases with larger statistical uncertainty owing to shorter averaging time).
}
\label{PIC9}
\end{figure*}
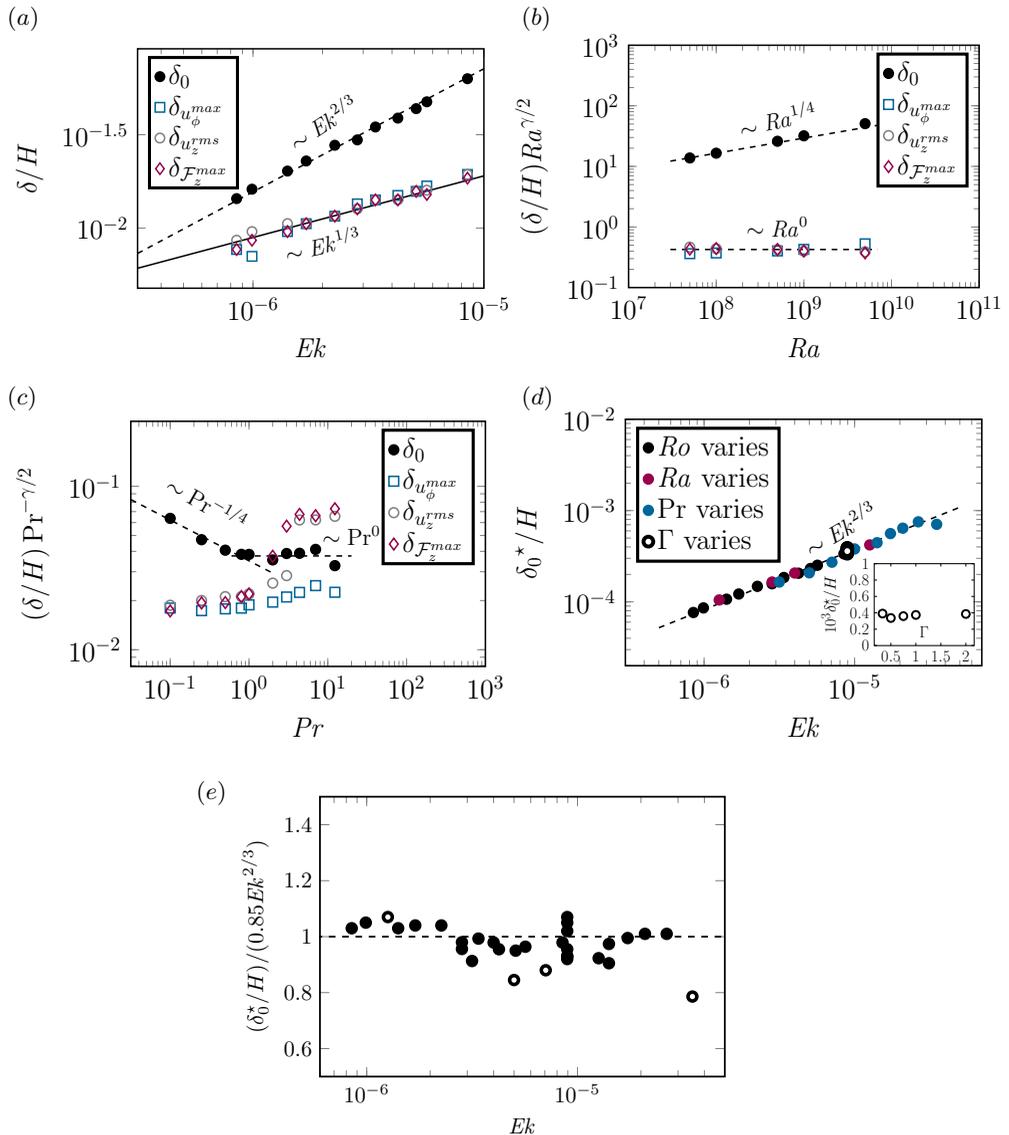

We next consider the quantitative dependence of the different layer widths on $\Ra$, $\Ek$ and $\Pr$, looking for a universal scaling form  $\delta/H \sim \Pran^\xi  \Ra^\beta \Ek^\gamma$.  In figure \ref{PIC9}a, the dependence of $\delta_0/H$ on $\Ek$ for $\Ra = 10^9$, $\Pran = 0.8$, and $ 2< 1/Ro <20 $ is shown to be consistent with a $\Ek^{2/3}$ scaling whereas the width based on other measures scale closely as $\Ek^{1/3}$, i.e., $\gamma$ takes on values of 2/3 and 1/3 for BZF width and velocity layer widths, respectively. (Because the statistical uncertainty in our reported exponents is of order 5--10\%, we report fractional scalings consistent with the data to within these uncertainties; they are not intended to denote exact results.)  
As mentioned in \cite{Zhang2020}, the BZF is characterized by bimodal temperature PDFs near the sidewall. This property was used in both DNS and experimental measurements to identify the BZF over a wide range of $\Ra$. Here, we conduct a more detailed analysis of the DNS data to explore how the width of the BZF changes with $\Ra$. We compute the width at fixed $\Ro = \Ra^{1/2} \Pran^{-1/2} \Ek$ so $\Ek=\Ro \Ra^{-1/2} \Pran^{1/2}$.
To determine the scaling with $\Ra$ at fixed $Ro=1/10$, we have that $\delta/H \sim \Ra^{\beta -\gamma/2}$. By multiplying by $\Ra^{\gamma/2}$ we obtain the scaling exponent $\beta$.  In figure \ref{PIC9}b, we plot $(\delta_0/H) \Ra^{1/3}$ and $(\delta/H) \Ra^{1/6}$ corresponding to $\gamma$ values of 2/3 and 1/3, respectively.  From this plot, we obtain values for $\beta$ of 1/4 and 0, respectively.  Similarly for the dependence on $\Pran$, we plot in figure \ref{PIC9}c the corrected quantities $(\delta/H)  \Pran^{\gamma/2}$ which yields $\delta_0/H$ scalings for $\xi$ of $-1/4$ for $\Pran < 1$ and 0 for $\Pran > 1$.  The other layer widths based on $u_\phi$, $u_z$ and $\mathcal{F}_z$ are independent of $\Pran$ for $\Pran < 1$ but do not collapse for $\Pran >1$.  The separation of the different widths for $\Pran >1$ suggests some interesting behaviour not captured by our scaling ansatz. Finally, we can collapse all the data for BZF width onto a single scaling curve by plotting in figure \ref{PIC9}d  $\delta_0^\star/H = \delta_0/H \left (\Pran^{\{1/4,\ 0\}} \Ra^{-1/4}\right )$ versus $\Ek$ (to compact the different scalings with $\Pran$ we denote them as $\Pran^{\{1/4,\ 0\}}$ for scaling with $\Pran < 1$ and $\Pran>1$, respectively) so that we can conclude that $\delta_0/H \sim \Pran^{\{-1/4,\ 0\}} \Ra^{1/4} \Ek^{2/3}$.  The results at one set of parameter values $\{\Ra, \Pran, \Ro \}$ are independent of $\Gamma$, see figure \ref{PIC9}d inset, which implies that $\delta_0/H \sim \Gamma^{0}$ (other dependences on $\Gamma$ are not ruled out for other parameter values although it is reasonable to assume it to be general in the absence of other data).
Thus, we plot in figure \ref{PIC9}d all the data with different $\Gamma$, $\Pran$, $\Ra$ and $\Ek$ to obtain scalings
\begin{eqnarray}
\delta_0/H &\approx& 0.85 \Gamma^{0} \Pran^{-1/4} \Ra^{1/4} \Ek^{2/3} \ \text{for}\ \Pran<1, \label{eq:small}\\
\delta_0/H &\approx& 0.85 \Gamma^{0} \Pran^{0} \Ra^{1/4} \Ek^{2/3} \ \text {for}\ \Pran>1. \label{eq:large}
\end{eqnarray}
We plot in figure \ref{PIC9}e the scaled BZF width 
$\left(\delta_0/H\right) / \left(0.85 \Gamma^{0} \Pran^{\{-1/4,\ 0\}} \Ra^{1/4} \Ek^{2/3}\right)$.  
One sees that the data scatter randomly within $\pm 10$\%, quite good agreement.

\begin{figure*}
\unitlength1truecm
\begin{picture}(13, 4)
\put(0, 0){\includegraphics[width=4.25cm]{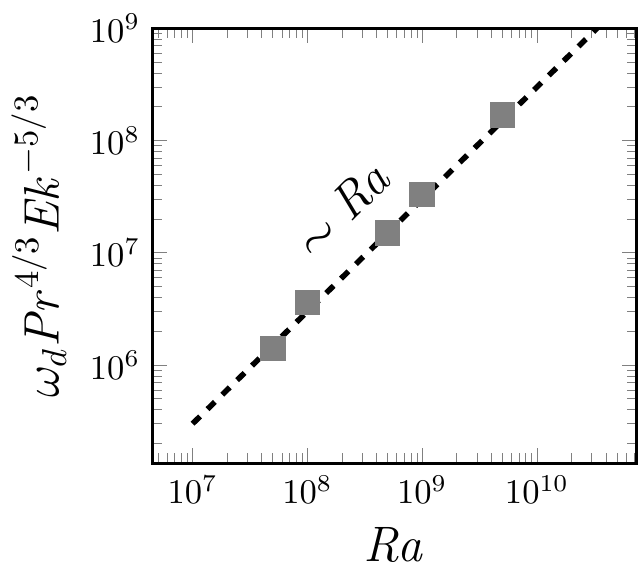}}
\put(4.5, 0){\includegraphics[width=4.4cm]{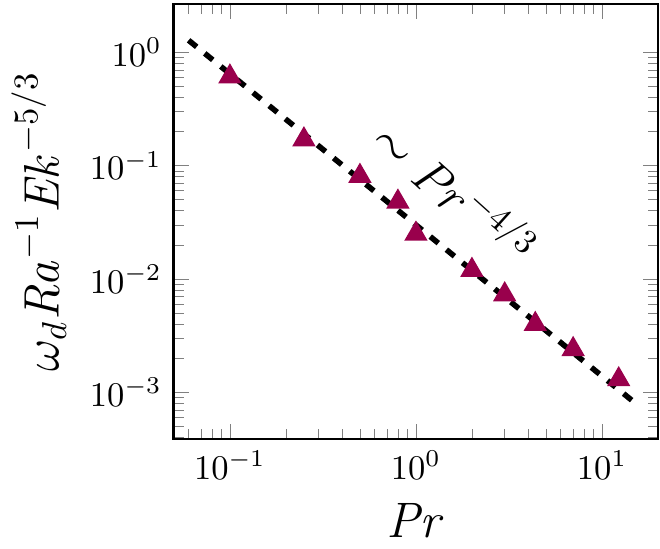}}
\put(9.2, 0){\includegraphics[width=4.5cm]{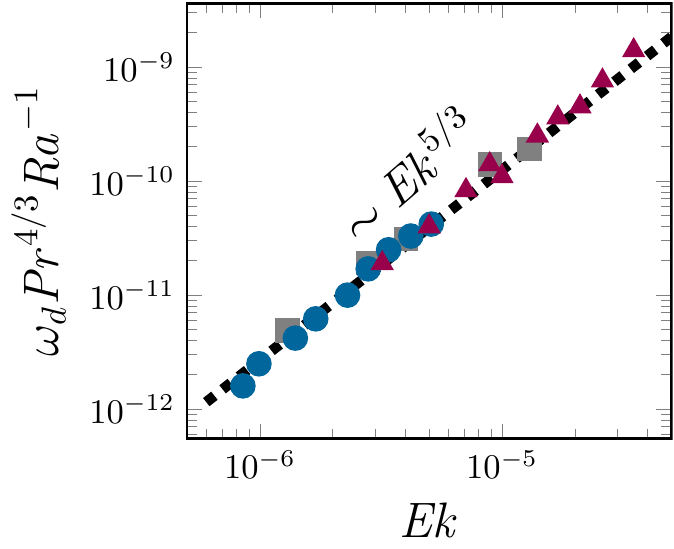}}
\put(-0.2, 3.7){$(a)$}
\put(4.3, 3.7){$(b)$}
\put(9.0, 3.7){$(c)$}
\end{picture}
\caption{
Scalings of $\omega_d$: (a) data scaled by $\Pran^{4/3} \Ek^{-5/3}$ showing $\Ra$ scaling, (b) scaled by $\Ra^{-1} \Ek^{-5/3}$ showing $\Pran^{-4/3}$ scaling, and (c) scaled by $\Pran^{4/3} \Ra^{-1}$ showing $\Ek^{5/3}$ scaling (cases at different $\Ra$ ({\tiny $\color{gray}{\blacksquare}$}), at different $\Pran$ ({\scriptsize $\color{gfred4}{\blacktriangle}$}), and at different $\Ro$ ({\large \color{gfblue4}{\textbullet}})).
}
\label{PIC10}
\end{figure*}
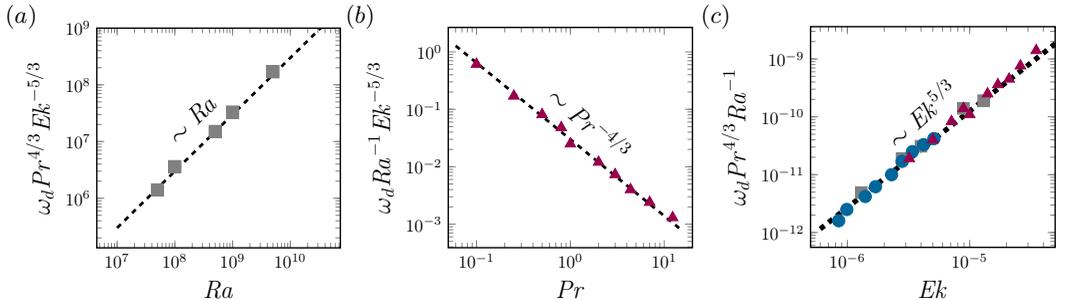

The BZF drifts anticyclonically, the same as the direction of traveling wall modes of rotating convection \citep{Zhong1991, Ecke1992, Kuo1993, Herrmann1993}. We plot in figure \ref{PIC10}a the drift frequency $\omega_d \equiv \omega/\Omega$ versus $\Ra$ showing scaling as $Ra$ and in figure \ref{PIC10}b versus $\Pran$ showing scaling as $\Pran^{-4/3}$(data in both are corrected for constant $\Ro$ conditions).  In figure \ref{PIC10}c, we scale out the dependence on $\Ra$ and $\Pran$, i.e., $\omega_d \Ra^{-1} \Pran^{4/3}$ and observe reasonable collapse with  $Ek^{5/3}$ scaling.   
From the cases listed in table \ref{TAB2}, we get the frequency scaling in terms of $\Ra$, $\Pran$, $\Gamma$, and $\Ek$ as 
\begin{equation}
\omega_d \approx 0.03 \Gamma^{0} \Pran^{-4/3} \Ra \Ek^{5/3}.
\label{eq:freq}
\end{equation}
The linear dependence on $\Ra$ is consistent with earlier results \citep{Horn2017,Wit2020,Favier2020} and suggests that there is a correspondence between the states we observe and the nonlinear manifestation of linear wall mode states.  The scalings we have determined for $\omega_d$ with $\Ek$ and $\Pr$ will be useful in making a more quantitative comparison with the wall mode hypothesis among data sets with different $\Ek$ and $\Pr$.  Such an analysis is beyond the scope of the present work and will be presented elsewhere. 
These scalings, of course, depend on the definition of the time unit.  Using the free-fall time or the vertical thermal diffusion time, respectively we obtain 
\begin{eqnarray}
\omega/\sqrt{\alpha g \Delta/H} &\approx& 0.015 \Gamma^{0} \Pran^{-5/6} \Ra^{1/2} \Ek^{2/3}\\
\omega/\left ( \kappa/H^2\right ) &\approx& 0.015 \Gamma^{0} \Pran^{-1/3} \Ra \Ek^{2/3},
\label{eq:freq1}
\end{eqnarray}
which both show the same $\Ek$ scaling as $\delta_0$, i.e., $\Ek^{2/3}$, see figure \ref{PIC9}a. 
For the three choices of time scale, the drift frequency decreases with increasing $\Pran$ for all $\Pran$ as opposed to the scaling of $\delta_0/H$ which has different scaling for small and large $\Pran$ (see figure \ref{PIC9}c and figure \ref{PIC12}b).

\begin{figure*}
\unitlength1truecm
\begin{picture}(12, 4.5)
\put(0, 0){\includegraphics[width=6.6cm]{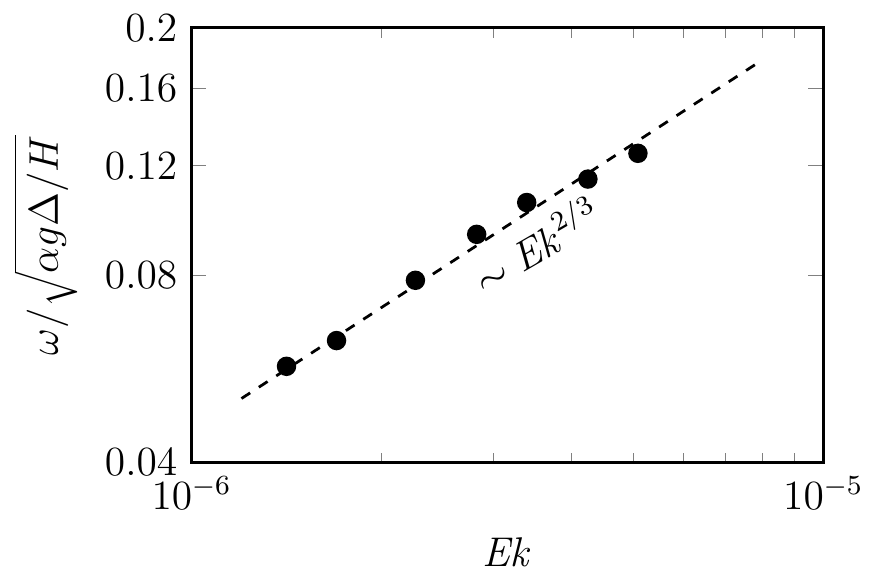}}
\put(6.6, 0){\includegraphics[width=6.6cm]{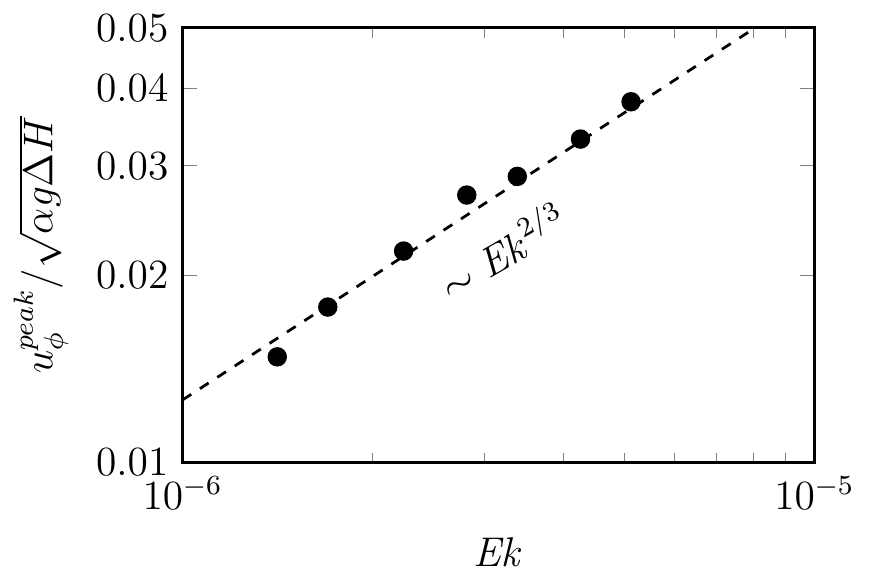}}
\put(0, 4.4){$(a)$}
\put(6.6, 4.4){$(b)$}
\end{picture}
\caption{For fixed $\Ra=10^{9}$ and rotation rates, $1/\Ro$=5.6, 6.7, 8.3, 10, 12.5, 16.7, 20: 
$(a)$ drift frequency $\omega$ of BZF,
$(b)$ maximum absolute value of $u_{\phi}$ near plates (mean value of two maxima).
Everywhere $\Pran=0.8$, $\Gamma=1/2$.
}
\label{PIC11}
\end{figure*}

\begin{figure*}
\center
\unitlength1truecm
\begin{picture}(12,6)
\put(-0.5, 0){\includegraphics[width=6.6cm]{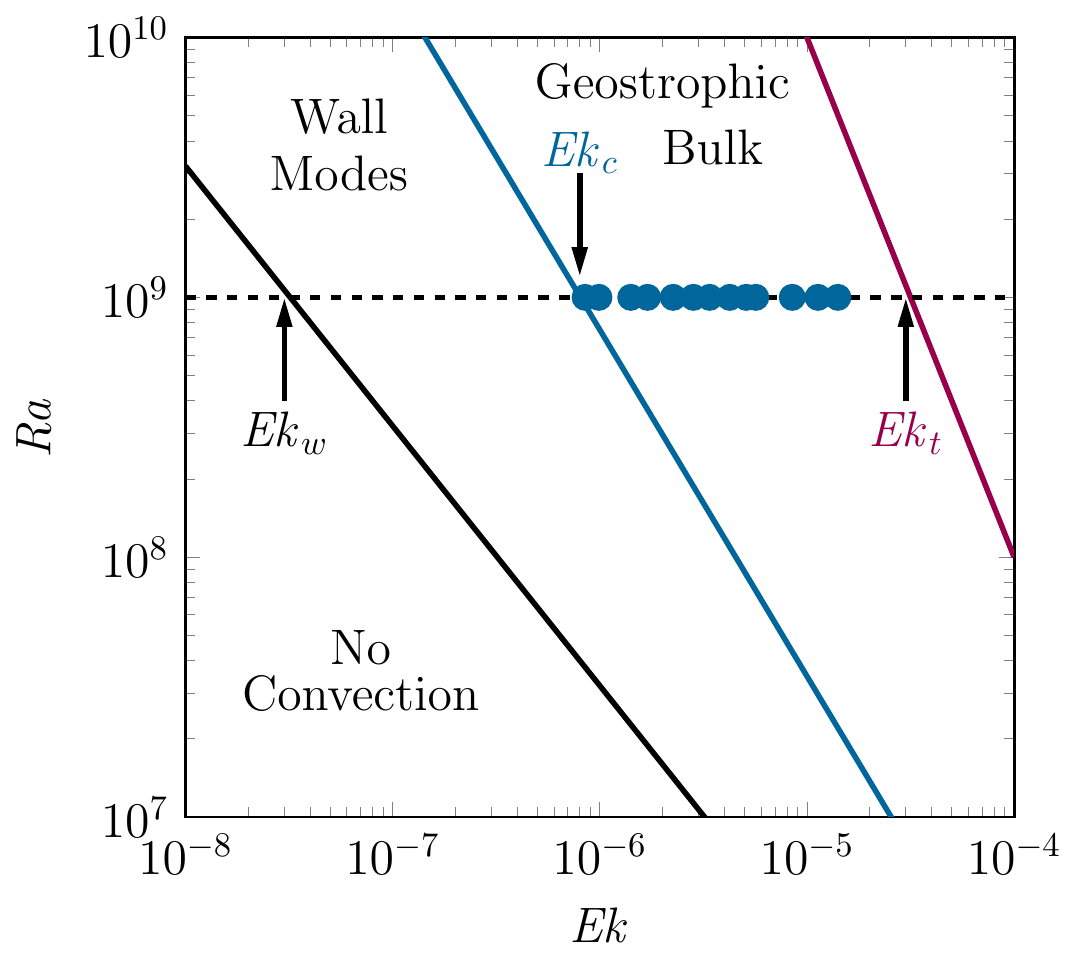}}
\put(6.1, 0){\includegraphics[width=6.8cm]{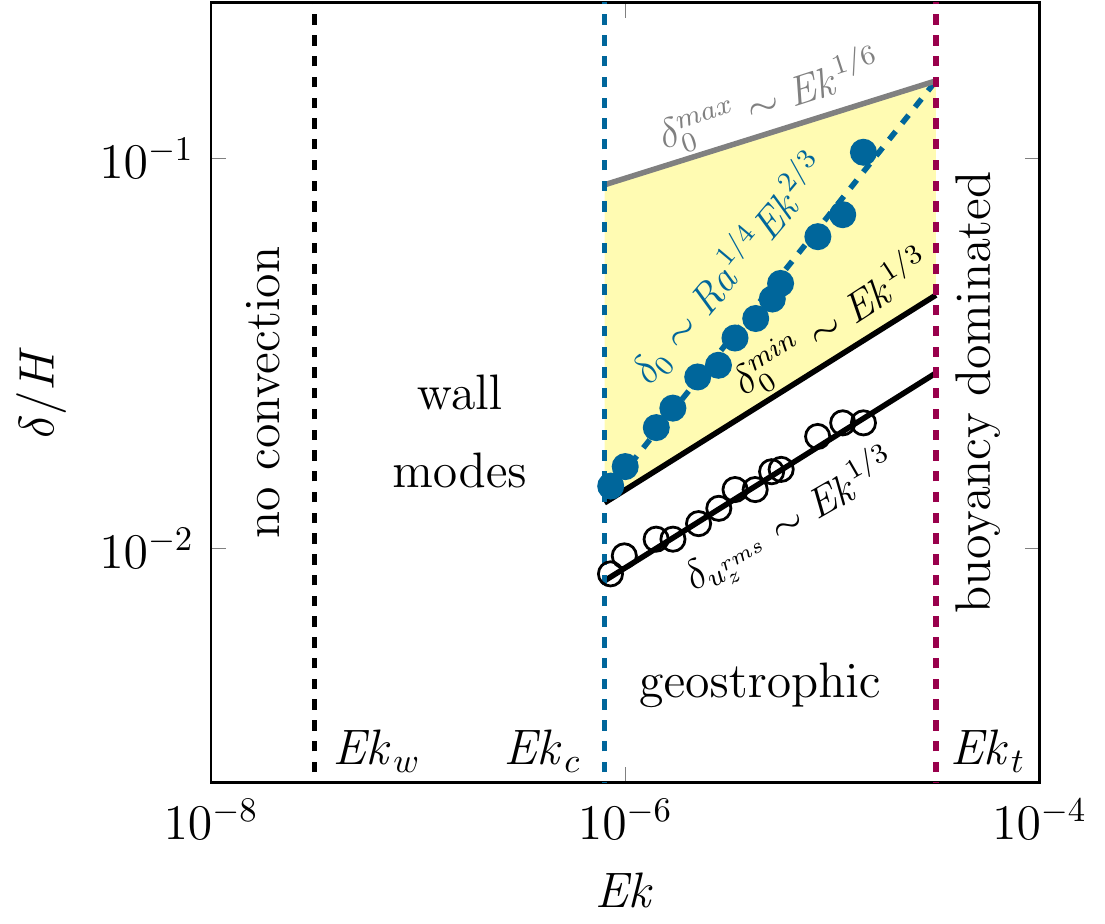}}
\put(-0.5, 6.2){$(a)$}
\put(6.3, 6.2){$(b)$}
\end{picture}
\caption{
(a) $\Ra - \Pran$ phase diagram.  Different rotating convective states are labeled.  Dashed horizontal line corresponds to $\Ra = 10^9$.  Critical $\Ek$ values $\Ek_w$, $\Ek_c$, and $\Ek_t$ are indicated. $\Ek$ data (solid circles) correspond to values in (b). (b) Widths of BZF $\delta_0$ and Stewartson $\sim \Ek^{1/3}$ layer $\delta_{u_z}^{\text{rms}}$ at $\Ra=10^9$ (DNS) vs $\Ek$.
Vertical lines (black, blue, red) are the critical Ekman numbers for onset of wall modes ($\Ek_w$), onset of bulk convection \citep{Chandrasekhar1961, Niiler1965}($\Ek_c$), and transition to rotation dominated regimes ($\Ek_t$) for $\Pran=0.8$, $\Ra=10^9$, $\Gamma=1/2$.
}
\label{PIC12}
\end{figure*}

As reported in \cite{Zhang2020} and shown here in figure \ref{PIC3}, the thermal structures drift anti-cyclonically, opposite to the azimuthal velocity which is cyclonic near the sidewall, as shown in figures \ref{PIC2}b-d. We show in figure \ref{PIC11}a that  the drift frequency decreases as rotation increases with a scaling $\Ek^{2/3}$. In figure \ref{PIC11}b, we show that the near-plate azimuthal velocity $u_\phi^{peak}$ is also anticyclonic and shows the same scaling behaviour with $\Ek$ (see figure \ref{PIC10}b) as the BZF width and drift frequency.  Based on this observation, we believe that the drift characteristics of the BZF are determined not only by the presence of the vertical wall but also by the near-plate region.

Finally, we consider the range of $\Ra$ and $\Ek$ in which the BZF is observed in this study.  There are three regions defined by the onset of wall-mode convection $\Ra_w \approx 32 \Ek^{-1}$, the onset of bulk convection $\Ra_c = A \Ek^{-4/3}$ , and the transition from geostrophic convection \citep{Grooms2010,Julien2012} to buoyancy dominated convection $\Ra_t = \Pran \Ro_t^2 \Ek^{-2}$ where $\Ro_t \approx 1$ (see Appendix figure \ref{PIC13}) is the transition Rossby number out of the geostrophic regime \citep{Julien1996,Liu2009,King2009,Weiss2011} as indicated in the $\Ra - \Ek$ phase diagram in figure \ref{PIC12}a.  Our data fall solely within the geostrophic regime of bulk convection but we include the other zones for context. According to \cite{Chandrasekhar1961} \citep[see also][]{Clune1993}, the critical Rayleigh number for the onset of convection is $\Ra_{c} \sim \Ek^{-4/3}$ with a prefactor $A$ that is weakly dependent on $\Ek$, in the range 6--8.7 \citep{Chandrasekhar1961,Niiler1965}; we use a value of 7.5 consistent with our range of $\Ek$. To illustrate one aspect of this range, we consider the BZF width $\delta_0/H$ versus $\Ek$ for $\Ra = 10^9$ in figure \ref{PIC12}b.
A path of constant $\Ra = 10^9$ (figure \ref{PIC12}a) yields $\Ek_w \approx 32 Ra^{-1} = 3.2 \times 10^{-8}$, $\Ek_c = (A Ra^{-1})^{3/4} = 8 \times 10^{-7}$, and $\Ek_t = \Ro_t \Pran^{1/2} Ra^{-1/2} = 2.8 \times 10^{-5}$.
Here the subscripts `w', `c' and `t' correspond, respectively, to the onset of wall-mode, bulk convection and transition from rotation to buoyancy dominated regime.
These values are indicated by vertical dashed lines in figure \ref{PIC12}b. 
Knowing the dependence of the critical $\Ra_{c}$ and $\Ek$ and using the relations (\ref{eq:small})(\ref{eq:large}), we can evaluate the smallest possible $\delta_{0}$ for any fixed $\Ek$, i.e., $\delta_0^{min} \sim\Ra_c^{1/4}\Ek^{2/3} \sim \Ek^{1/3}$ (see $\delta_0^{min}$ in figure \ref{PIC12}b).
Connecting these onset points, we obtain the black line in the diagram, which is parallel to the Stewartson ``1/3" layer scaling. The gap between these two black solid lines depends slightly on A but the ratio of the BZF width to the Stewartson layer width is constant (based on $\delta^{rms}_{u_z}$) at the onset of convection (the fixed gap).
Thus, although the BZF width decreases faster than the Stewartson layer as rotation increases, there is no crossing of the BZF boundary and the boundary of the Stewartson layer at extreme fast rotation because bulk convection ceases before they can cross.  Note that all the data considered here fall within the geostrophic range of rotating convection; what happens in the wall-mode region is not addressed.
 
The other bound on the BZF scaling depends on when rotation becomes significant. 
An estimate is made based on a plot of $\Nu/\Nu_{0}$ versus $\Pran \Ro^2$ for our DNS and for experimental data from \cite{Wedi2020}, see Appendix figure \ref{PIC13}, where the data for $\Ra$ from $10^8$ to $10^{14}$ merge onto a single curve. 
Here $\Nu_0$ is the Nusselt number in non-rotating case.
Using an empirical estimate $\Ro_t \approx 1$ for the onset of the rotation dominated regimes, i.e., the geostrophic regime, we get an estimate for the largest possible $\delta_0$, for any $\Ek$ (grey line in figure \ref{PIC12}, that is, $\delta_0^{max} \propto \Ra_t^{1/4}\Ek^{2/3} \propto \Ro_t^{1/2}\Ek^{1/6} \approx \Ek^{1/6}$).
($\Pran\Ro_t^2 \approx 1$ is the onset in figure \ref{PIC13}, but in experiment $\Pran$ varies from 0.7 to 0.9 and in DNS $\Pran=0.8$, thus here we take $\Pran=1$ which gives $\Ro_t \approx 1$ for simplicity.)

It is remarkable that the BZF regime is confined by these two critical lines ($\sim\Ek^{1/6}$ and $\sim\Ek^{1/3}$) and the range confined in between gets broader for higher $\Ra$. In other words, at low $\Ra$, the BZF is only observed over a small range of rotation rates.
At large $\Ra$, the BZF exists over a much broader range of rotation rates \citep{Zhang2020,Wedi2020}.
For any fixed $\Ra$, the BZF exists in a certain $\Ek$-range which is determined by the grey and black lines in figure \ref{PIC12}b and the BZF width changes as $\delta_0\sim\Ek^{2/3}$ over that range. How the BZF contributes to the heat transport relative to the contribution of the laterally unbounded system in the geostrophic regime remains an open question.  Further, the connection between the BZF and linear wall modes requires additional work to understand the relationship between the two convective states.

\section{Conclusion}
The BZF is found to be an important flow structure in rapidly rotating turbulent Rayleigh--B\'enard convection in the geostrophic regime and is robust over considerable ranges of $\Ra$, $\Ek$, $\Pran$ and $\Gamma$. 
The main structure, drift of plume pairs, is found to be a $m=2\Gamma$-mode for the choices of $\Gamma = 1/2, 1, 2$; additional values of $\Gamma = 1/3, 3/4$ yield $m=1$ suggesting mode 1 for $\Gamma \lesssim 1$.  In addition, the BZF carries a large portion of the total heat; 
its contribution to the total heat transport is about $60\%$ of the heat transport at fast rotation, $\Ro < 0.1$, and for $\Pran <1$. For $\Pran > 1$, the BZF heat transport contribution drops to about $35\%$. Understanding this important contribution to the heat transport is essential in analyzing experiments in rotating convection in the geostrophic regime. 

The scaling of the BZF width $\delta_0$ depends on $\Pran$, $\Ra$ and $\Ek$ as $\delta_0/H \sim \Gamma^{0} \Pr^{\{-1/4,\ 0\}} \Ra^{1/4} \Ek^{2/3}$ ($\Pran^{-1/4}$ for small-to-moderate $\Pran$ and independent of $\Pran$ for large $\Pran$). The universal scaling of the BZF and the sidewall boundary layers is very clean for $\Pran < 1$ but the BZF is less coherent for $\Pran >1$ and the sidewall boundary layer widths behave differently for those conditions.  Further, the sharp decrease in the BZF heat transport contribution similarly marks a transition to a perhaps more complex BZF state for $\Pran > 1$. The drift frequency of the BZF shows scaling $\omega/\Omega \sim \Gamma^{0} \Pr^{-4/3} \Ra \Ek^{5/3}$, indicating that the drift frequency decreases significantly as $\Pran$ increases, is proportional to $\Ra$, and decreases rapidly with increasing rotation (decreasing $\Ek$). Interestingly, $\omega$ seems to be more robust than $\delta$ with respect to changes in $\Pran$. Finally, the BZF shares qualitative and some quantitative characteristics with linear wall modes and establishing the connection between these two states will be an important area of future research.

\section{Acknowledgement}
The authors would like to thank Eberhard Bodenschatz, Detlef Lohse, Marcel Wedi and Stephan Weiss for the fruitful discussions, cooperation and support.
The authors acknowledge Leibniz Supercomputing Centre (LRZ) for providing computing time. 

\section{Funding}
This work was supported by the Deutsche Forschungsgemeinschaft (X.Z. and O.S., grant number Sh405/8, Sh405/7 (SPP 1881 Turbulent ``Superstructures"));  and the Los Alamos National Laboratory LDRD program under the auspices of the U.S. Department of Energy (R.E.E.).

\section{Declaration of Interests}
The authors report no conflict of interest.

\section{Appendix}

We tabulate here a full characterization of the parameters in the DNS (see table \ref{TAB2}) and compare our results for $Nu$ with experimental data from \cite{Wedi2020}. The excellent agreement is a strong indication that our DNS are fully resolved. We also include the resulting numerical values of the BZF drift frequency for different choices of time scale (see table \ref{TAB3}), namely $\omega_\kappa$, $\omega_{ff}$, and $\omega_d$ for different parameters values $\Gamma$, $\Pran$, $\Ra$, $\Ek$, $1/\Ro$.

\begin{figure*}
\center
\includegraphics[width=10cm]{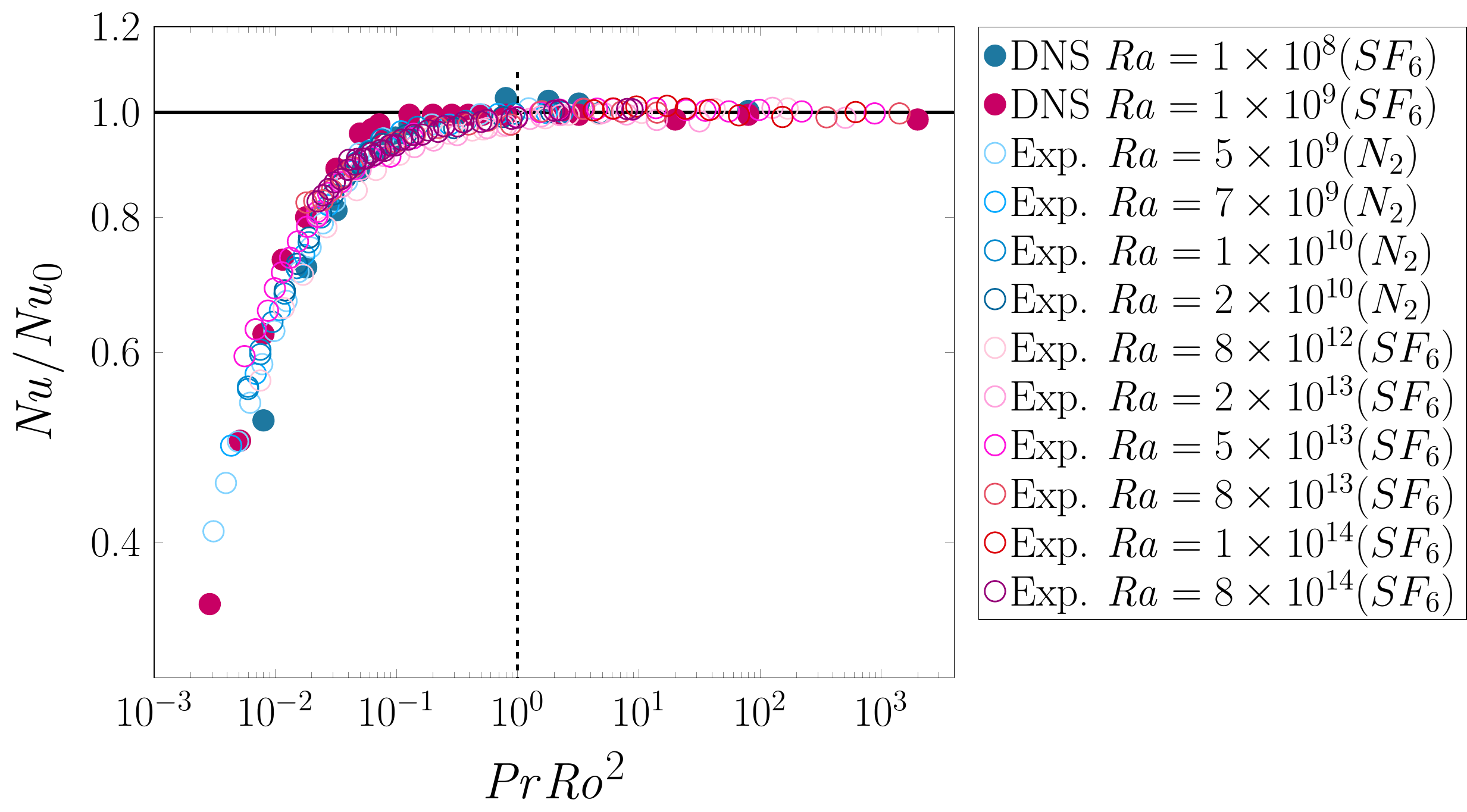}
\caption{
Double logarithmic scale plot of $\Nu/\Nu_0$ vs.
$\Pran\Ro^2$.  Horizontal line indicates $\Nu/\Nu_0 = 1$; vertical line indicates the value $\Ro_t$, i.e. a transition between buoyancy dominated convection at larger $\Ro$ ($\Nu \approx \Nu_0$) and the rotation dominated regime at smaller $\Ro$ ($\Nu < \Nu_0$) .  Experimental data are from \cite{Wedi2020}.
}
\label{PIC13}
\end{figure*}

\begin{landscape}
\begin{table*}
\begin{center}
 \scriptsize
        \begin{tabular}[t]{lccrrrrrcccrrrc}
        \toprule
            $\Gamma$& $\qquad\Pran\qquad$ & $\quad\Ra\quad$ & $\quad1/\Ro\;$ & $\quad t_{\text{avg}}/\tau_{ff}$ & $\ N_r\ $ & $\ N_\phi\ $ & $\ N_z\ $& $\ \mathcal{N}_{\text{th}}\ $&$\
            \mathcal{N}_{\text{u}}\ $&$\ \mathcal{N}^{\text{sw}}_{\text{u}}\ $ & $\delta_\text{th}/H\qquad$ & $\delta_\text{u}/H\qquad$ & $\delta^{\text{sw}}_\text{u}/H\qquad$ & max($h/\eta_k$)\\ \hline
             1/2 & 0.8& $5.0 \times 10^7$& 10  & 440 & 90 & 256 & 380 & 43 & 49  & 16  &$3.5\times10^{-2}$ & $4.5\times10^{-2}$ & $2.3\times10^{-2}$ & 0.63\\
                   &      & $1.0  \times 10^8$& 10  &  500  & 100 & 256 & 380 & 39  & 50   & 17 & $3.0\times10^{-2}$ & $4.9\times10^{-2}$  & $2.0\times10^{-2}$ & 0.79 \\
                   &      & $5.0  \times 10^8$& 10  &   370   & 128 & 320 & 620 & 46  & 89   & 17 & $1.6\times10^{-2}$ & $5.4\times10^{-2}$  & $1.4\times10^{-2}$ & 0.67\\
                   &      & $1.0  \times 10^9$& 10  &   1130 & 192 & 512 & 820 & 37  & 101 & 18 & $1.2\times10^{-2}$ & $5.3\times10^{-2}$  & $1.3\times10^{-2}$ & 0.85\\
                   &      & $5.0  \times 10^9$& 10  &   200   & 256 & 620 & 860 &  39 & 118  & 27& $7.3\times10^{-3}$ & $5.1\times10^{-2}$  & $9.4\times10^{-3}$ & 1.24  \\\hline
             1/2 & 0.1 & $1.0  \times 10^8$& 10  &  750 & 256 & 256 & 512 & 70 & 60 & 33 &$7.3\times10^{-2}$ & $5.7\times10^{-2}$ & $1.2\times10^{-2}$ & 1.10\\
             	  & 0.25  &    			  & 10  &  80 & 100 & 256 & 380 & 39 & 50  & 17 &$5.1\times10^{-2}$ & $5.2\times10^{-2}$ & $1.9\times10^{-2}$ & 0.77\\
             	  & 0.5  &    			  & 10  &  250 & 100 & 256 & 380 & 39 & 50  & 17 &$4.2\times10^{-2}$ & $4.9\times10^{-2}$ & $1.7\times10^{-2}$ & 0.92\\
                   & 0.8  &    			  & 10  &  500 & 100 & 256 & 380 & 39 & 50  & 17 &$3.0\times10^{-2}$ & $4.9\times10^{-2}$ & $2.0\times10^{-2}$ & 0.79\\
                   & 1     &    			  & 10  &  450 & 100 & 256 & 380 & 37 & 50  & 17 &$2.8\times10^{-2}$ & $4.8\times10^{-2}$ & $2.0\times10^{-2}$ & 0.72\\

                   & 2     &    			  & 10  &  500 & 100 & 256 & 380 & 29 & 53  & 22 &$1.8\times10^{-2}$ & $5.3\times10^{-2}$ & $2.9\times10^{-2}$ & 0.80\\
                   & 3     &    			  & 10  &  600 & 100 & 256 & 380 & 26 & 53  & 35 &$1.5\times10^{-2}$ & $5.2\times10^{-2}$ & $3.4\times10^{-2}$ & 0.83\\
                   & 4.38&    			  & 10  &  430 & 100 & 256 & 380 & 24 & 50  & 16 &$1.3\times10^{-2}$ & $4.8\times10^{-2}$ & $1.6\times10^{-1}$ & 0.86\\
                   & 7     &    			  & 10  & 540& 100 & 256 & 380 &  24   & 46  & 49 & $1.3\times10^{-2}$  &  $4.2\times10^{-2}$ & $8.9\times10^{-2}$ & 0.88\\
                   & 12.3&    		 	  & 10  & 420 & 100 & 256 & 380 &  23  & 41  & 54 & $1.2\times10^{-2}$  & $3.3\times10^{-2}$ & $1.0\times10^{-1}$ & 0.88\\\hline
             1/3 & 0.8 & $1.0 \times 10^8$ & 10  &  810 & 96 & 256 & 320 & 28 & 34  & 22  &$2.8\times10^{-2}$ & $3.8\times10^{-1}$ & $2.0\times10^{-2}$ & 0.59\\
             1/2 &       &                              & 10  &  500 & 100 & 256 & 380 & 39 & 50  & 17  &$3.0\times10^{-2}$ & $4.9\times10^{-2}$ & $2.0\times10^{-2}$ & 0.79\\
             3/4 &       &    			  & 10  &  830 & 128 & 256 & 380 & 38 & 53  & 17 &$3.4\times10^{-2}$ & $5.8\times10^{-1}$ & $2.0\times10^{-2}$ & 0.91\\
             1 &       &    			          & 10  &  1480 & 180 & 320 & 320 & 22 & 36  & 15 &$3.5\times10^{-2}$ & $6.5\times10^{-2}$ & $2.0\times10^{-2}$ & 0.90\\
             2 &       &    			          & 10  &  1500 & 256 & 320 & 320 & 25 & 46  & 14 &$4.0\times10^{-2}$ & $9.0\times10^{-2}$  & $2.1\times10^{-2}$ & 1.24\\ \hline
             1/2 & 0.8 & $1.0  \times 10^9$& 0.5  &  965  & 192 & 512 & 768 & 14 & 78 & 42 & $7.8\times10^{-3}$ & $5.6\times10^{-2}$ & $3.4\times10^{-2}$ & 0.94\\
                   &       &                              & 2     &  1020 & 192 & 512 & 768 & 14 & 87 & 28 & $7.8\times10^{-3}$ & $6.5\times10^{-2}$ & $2.1\times10^{-2}$ & 0.94\\
                   &       &                              & 2.5  &  1410 & 192 & 512 & 768 & 14 & 88 & 28 & $7.7\times10^{-3}$ & $6.6\times10^{-2}$ & $2.1\times10^{-2}$ & 0.94\\
                   &       &                              & 3.3  &  1400 & 192 & 512 & 768 & 15 & 90 & 26 & $8.0\times10^{-3}$ & $6.8\times10^{-2}$ & $2.0\times10^{-2}$ & 0.94\\
             	  &       &    			  & 5     &  1630 & 192 & 512 & 768 & 16 & 88  & 22 & $8.8\times10^{-3}$ & $6.5\times10^{-2}$ & $1.6\times10^{-2}$ & 0.92\\
             	  &       &    			  & 5.6  &  480   & 180 & 280 & 680 & 34 &105 & 25 & $9.0\times10^{-3}$ & $6.5\times10^{-2}$ & $1.6\times10^{-2}$ & 1.13\\
                   &       &    			  & 6.7  &  480   & 180 & 280 & 680 & 35 &102 & 23 & $9.8\times10^{-3}$ & $6.3\times10^{-2}$ & $1.4\times10^{-2}$ & 1.10\\
                   &       &    			  & 8.3  &  500 & 180 & 280 & 680 & 38 & 99  & 23 & $1.1\times10^{-2}$ & $5.8\times10^{-2}$ & $1.4\times10^{-2}$ & 1.07\\

                   &        &    			  & 10    & 1130 & 192 & 512 & 820 & 37 &101 & 18 & $1.2\times10^{-2}$  & $5.3\times10^{-2}$ & $1.3\times10^{-2}$ & 0.85\\
                   &        &    			  & 12.5 &  400  & 128 & 320 & 620 & 45 & 83  & 15 & $1.5\times10^{-2}$  & $4.8\times10^{-2}$ & $1.2\times10^{-2}$ & 1.20\\
                   &        &    			  & 16.7 &  400  & 128 & 320 & 620 & 54 & 73  & 13 & $2.1\times10^{-2}$  & $3.8\times10^{-2}$ & $1.1\times10^{-2}$ & 1.10\\
                   &        &    			  & 20    &  400  & 128 & 320 & 620 & 63 & 65  & 12 & $2.8\times10^{-2}$  & $3.0\times10^{-2}$ & $9.5\times10^{-3}$ & 1.03\\
                   &        &    		 	  & 28.6 &  40    & 128 & 320 & 620 & 83 & 49  & 12 & $4.8\times10^{-2}$  & $1.8\times10^{-2}$ & $9.5\times10^{-3}$ & 0.90\\
                   &        &    			  & 33.3 &  37    & 128 & 320 & 620 & 86 & 48  & 11 & $5.0\times10^{-2}$  & $1.7\times10^{-2}$ & $8.5\times10^{-3}$ & 0.89\\
      \bottomrule
        \end{tabular}
\end{center} 
\caption{Details of the DNS including the
            time of statistical averaging, $t_{\text{avg}}$, normalised with
            the free-fall time $\tau_{ff}$; number of nodes $N_r$, $N_\phi$,
            $N_z$ in the directions $r$, $\phi$ and $z$, respectively;
            the number of the nodes within the thermal BL
            $\mathcal{N}_{\text{th}}$ (near the plates), within the viscous BL
            $\mathcal{N}_{\text{u}}$ (near the plates), and within the viscous BL
            $\mathcal{N}^{\text{sw}}_{\text{u}}$ (near the sidewall);
            the relative thickness of the viscous BL $\delta_u/H$
            and the thermal BL near plates $\delta_\text{th}/H$, the viscous BL near the sidewall $\delta^{\text{sw}}_\text{u}/H$, and the
maximal value of the ratio of the mesh size to the mean Kolmogorov microscale, max($h/\eta_k$).
 }
\label{TAB2}
\end{table*}
\end{landscape}

\begin{table}
\begin{center}
\begin{tabular}[t]{lcccccccc}
\toprule
$\Gamma$ & $\Pran$ & $\quad\Ra\quad$  & $\quad\Ek\quad$ & $1/\Ro$ & $\omega_\kappa$ & $\omega_{ff}$ & $\omega_d$ & $\omega_d^*$\\
\hline
1/2             & 0.8   & $5.0 \times 10^7$ 	    & $1.3 \times 10^{-5}$     & 10  &  $6.3 \times 10^{1}$  & $1.0 \times 10^{-2}$  & $1.3 \times 10^{-2}$  & $1.9 \times 10^{-10}$\\
                  &         & $1.0  \times 10^8$         & $8.9 \times 10^{-6}$     &  10  & $1.3 \times 10^{2}$  & $1.5 \times 10^{-2}$  & $1.8 \times 10^{-2}$  & $1.4 \times 10^{-10}$\\
                  &         & $5.0  \times 10^8$         & $4.0 \times 10^{-6}$     &  10  &  $3.3 \times 10^{2}$ & $1.7 \times 10^{-2}$  & $2.1 \times 10^{-2}$  & $3.1 \times 10^{-11}$\\
                  &         & $1.0  \times 10^9$         & $2.8 \times 10^{-6}$     &  10  &  $5.7 \times 10^{2}$ & $2.0 \times 10^{-2}$  & $2.5 \times 10^{-2}$  & $1.9 \times 10^{-11}$\\
                  &         & $5.0  \times 10^9$         & $1.3 \times 10^{-6}$     & 10  &  $1.7 \times 10^{3}$  & $2.6 \times 10^{-2}$  & $3.3 \times 10^{-2}$  & $4.9 \times 10^{-12}$\\
\hline
 1/2   & 0.1   & $1.0  \times 10^8$                      & $3.2 \times 10^{-6}$     & 10  &  $1.1 \times 10^{2}$  & $3.3 \times 10^{-2}$  & $4.2 \times 10^{-2}$  & $1.9 \times 10^{-11}$\\
         & 0.25  &                                                   & $5.0 \times 10^{-6}$     & 10  &  $1.0 \times 10^{2}$  & $2.0 \times 10^{-2}$  & $2.5 \times 10^{-2}$  & $4.0 \times 10^{-11}$\\
         & 0.5    &                                                   & $7.1 \times 10^{-6}$     & 10  &  $1.2 \times 10^{2}$  & $1.7 \times 10^{-2}$  & $2.1 \times 10^{-2}$  & $8.3 \times 10^{-11}$\\
         & 0.8    &                                                   & $8.9 \times 10^{-5}$     & 10  &  $1.3 \times 10^{2}$  & $1.5 \times 10^{-2}$  & $1.8 \times 10^{-2}$  & $1.4 \times 10^{-10}$\\
         & 1.0    &                                                   & $1.0 \times 10^{-5}$     & 10  &  $9.1 \times 10^{1}$  & $9.1 \times 10^{-3}$  & $1.1 \times 10^{-2}$  & $1.1 \times 10^{-10}$\\
         & 2.0    &                                                   & $1.4 \times 10^{-5}$     & 10  &  $1.1 \times 10^{2}$  & $8.0 \times 10^{-3}$  & $1.0 \times 10^{-2}$  & $2.5 \times 10^{-10}$\\
         & 3.0    &                                                   & $1.7 \times 10^{-5}$     & 10  &  $1.2 \times 10^{2}$  & $6.7 \times 10^{-3}$  & $8.4 \times 10^{-3}$  & $3.6 \times 10^{-10}$\\
         & 4.38  &                                                   & $2.1 \times 10^{-5}$     & 10  &  $1.0 \times 10^{2}$  & $5.0 \times 10^{-3}$  & $6.3 \times 10^{-3}$  & $4.5 \times 10^{-10}$\\
         & 7.0    &                                                   & $2.6 \times 10^{-5}$     & 10  &  $1.2 \times 10^{2}$  & $4.5 \times 10^{-3}$  & $5.7 \times 10^{-3}$  & $7.6 \times 10^{-10}$\\
         & 12.3  &                                                   & $3.5 \times 10^{-5}$     & 10  &  $1.4 \times 10^{2}$  & $4.0 \times 10^{-3}$  & $5.0 \times 10^{-3}$  & $1.4 \times 10^{-9}$\\
\hline
 1/2 	& 0.8 & $1.0 \times 10^8$ & $8.9 \times 10^{-6}$	& 10  &  $1.3 \times 10^{2}$  & $1.5 \times 10^{-2}$  & $1.8 \times 10^{-2}$  & $1.4 \times 10^{-10}$\\
 1 	&       &    				&	                                  & 10  &  $1.4 \times 10^{2}$  & $1.6 \times 10^{-2}$  & $2.0 \times 10^{-2}$  & $1.5 \times 10^{-10}$\\
 2 	&       &    				&	                                  & 10  &  $1.4 \times 10^{2}$  & $1.6 \times 10^{-2}$  & $2.0 \times 10^{-2}$  & $1.5\times 10^{-10}$\\
 \hline
             1/2 & 0.8 & $1.0  \times 10^9$	& $5.1 \times 10^{-6}$	& 5.6  &  $7.1 \times 10^{2}$  & $2.5 \times 10^{-2}$  & $5.7 \times 10^{-2}$  & $4.2 \times 10^{-11}$\\
                   &       &    			  	& $4.2 \times 10^{-6}$	& 6.7  &  $6.6 \times 10^{2}$  & $2.3 \times 10^{-2}$  & $4.4 \times 10^{-2}$  & $3.3 \times 10^{-11}$\\
                   &       &    			  	& $3.4 \times 10^{-6}$	& 8.3  &  $6.3 \times 10^{2}$  & $2.2 \times 10^{-2}$  & $3.4 \times 10^{-2}$  & $2.5 \times 10^{-11}$\\
                   &        &    			 	& $2.8 \times 10^{-6}$	& 10   &  $5.7 \times 10^{2}$  & $2.0 \times 10^{-2}$  & $2.5 \times 10^{-2}$  & $1.9 \times 10^{-11}$\\
                   &        &    			  	& $2.3 \times 10^{-6}$	& 12.5 &  $3.8 \times 10^{2}$  & $1.3 \times 10^{-2}$  & $1.3 \times 10^{-2}$  & $1.0 \times 10^{-11}$\\
                   &        &    			  	& $1.7 \times 10^{-6}$	& 16.7 &  $3.1 \times 10^{2}$  & $1.1 \times 10^{-2}$  & $8.4 \times 10^{-3}$  & $6.2 \times 10^{-12}$\\
                   &        &    			  	& $1.4 \times 10^{-6}$	& 20    &  $2.6 \times 10^{2}$  & $9.1 \times 10^{-3}$  & $5.7 \times 10^{-3}$  & $4.2 \times 10^{-12}$\\
                   &        &    		 	  	& $9.9 \times 10^{-7}$	& 28.6 &  $2.2 \times 10^{2}$  & $7.7 \times 10^{-3}$  & $3.4 \times 10^{-3}$  & $2.5 \times 10^{-13}$\\
                   &        &    			  	& $8.5 \times 10^{-7}$	& 33.3 &  $1.7 \times 10^{2}$  & $5.9 \times 10^{-3}$  & $2.2 \times 10^{-3}$  & $1.6 \times 10^{-13}$\\
\bottomrule
\end{tabular}
\caption{$\Gamma$, $\Ra$, $\Pran$, $\Ek$, $\Ro^{-1}$, $\omega_\kappa = \omega H^2/\kappa$, $\omega_{ff} = \omega \left ( H/\left ( g \alpha \Delta \right)\right)^{1/2}$, and   $\omega_d = \omega/\Omega$, $\omega_d^* = \omega_d \Pran^{4/3} \Ra^{-1}$. (For $\Gamma = 1/3, 3/4$, there is insufficient data to determine $\omega$.)}
\label{TAB3}
\end{center}
\end{table}
\FloatBarrier

\bibliographystyle{jfm}

\end{document}